\documentclass[acmtog,nonacm]{acmart}

\usepackage{booktabs} 

\acmPrice{15.00}

\setcitestyle{square}

\newif\ifmarkups
\markupsfalse

\usepackage[T1]{fontenc}

\usepackage{microtype}

\usepackage{amsmath}
\newcommand{\eqn}{ 
  \refstepcounter{equation}
  \eqno \hbox{{\normalfont \normalcolor (\theequation) }}
}
\newcommand{\mbf}[1]{\boldsymbol{\mathbf{#1}}} 

\usepackage[utf8]{inputenc}
\usepackage{alphabeta}
\usepackage{textalpha}

\usepackage{gensymb}
\DeclareUnicodeCharacter{00B0}{\degree}

\usepackage{graphicx} 
\usepackage{subcaption} 
\usepackage{wrapfig} 
\usepackage{booktabs} 
\usepackage{xcolor}
\usepackage{enumitem} 

\graphicspath{{img/}}

\newcommand{\user}[2]{\textsf{#2}} 

\ifmarkups
  \usepackage{pdfcomment} 
  \newcommand{\rev}[2]{{\color{red}{#1}}}
  \newcommand{\todo}[1]{{\color{blue}ToDo: *** {#1} ***}}
  \usepackage[nolabel]{showlabels,rotating}
  
\else
  \usepackage[final]{pdfcomment} 
  \newcommand{\todo}[1]{}
  \newcommand{\rev}[2]{#1}
\fi

\begin{document}

\title{ADD: Analytically Differentiable Dynamics for Multi-Body Systems with Frictional Contact}

\author{Geilinger, M.}
\affiliation{%
  \department{CRL}
  \institution{ETH Zürich}}

\author{Hahn, D.}
\affiliation{%
  \department{CRL}
  \institution{ETH Zürich}}

\author{Zender, J.}
\affiliation{%
  \department{LIGUM}
  \institution{Université de Montréal}}

\author{Bächer, M.}
\affiliation{%
  \institution{Disney Research Zürich}}

\author{Thomaszewski, B.}
\affiliation{%
  \department{CRL}
  \institution{ETH Zürich}}
\affiliation{\department{LIGUM}
  \institution{Université de Montréal}}
  
\author{Coros, S.}
\affiliation{%
  \department{CRL}
  \institution{ETH Zürich}}

\renewcommand{\shortauthors}{Geilinger, Hahn, Zender, Bächer, Thomaszewski, and Coros}

\begin{abstract}

We present a differentiable dynamics solver that is able to handle frictional contact for rigid and deformable objects within a unified framework. Through a principled mollification of normal and tangential contact forces, our method circumvents the main difficulties inherent to the non-smooth nature of frictional contact. We combine this new contact model with fully-implicit time integration to obtain a robust and efficient dynamics solver that is analytically differentiable. In conjunction with adjoint sensitivity analysis, our formulation enables gradient-based optimization with adaptive trade-offs between simulation accuracy and smoothness of objective function landscapes. We thoroughly analyse our approach on a set of simulation examples involving rigid bodies, visco-elastic materials, and coupled multi-body systems. We furthermore showcase applications of our differentiable simulator to parameter estimation for deformable objects, motion planning for robotic manipulation, trajectory optimization for compliant walking robots, as well as efficient self-supervised learning of control policies.
\end{abstract}




\begin{teaserfigure}
\centering
\def\svgwidth{\textwidth} 
\begingroup%
  \makeatletter%
  \providecommand\color[2][]{%
    \errmessage{(Inkscape) Color is used for the text in Inkscape, but the package 'color.sty' is not loaded}%
    \renewcommand\color[2][]{}%
  }%
  \providecommand\transparent[1]{%
    \errmessage{(Inkscape) Transparency is used (non-zero) for the text in Inkscape, but the package 'transparent.sty' is not loaded}%
    \renewcommand\transparent[1]{}%
  }%
  \providecommand\rotatebox[2]{#2}%
  \newcommand*\fsize{\dimexpr\f@size pt\relax}%
  \newcommand*\lineheight[1]{\fontsize{\fsize}{#1\fsize}\selectfont}%
  \ifx\svgwidth\undefined%
    \setlength{\unitlength}{5070.84155273bp}%
    \ifx\svgscale\undefined%
      \relax%
    \else%
      \setlength{\unitlength}{\unitlength * \real{\svgscale}}%
    \fi%
  \else%
    \setlength{\unitlength}{\svgwidth}%
  \fi%
  \global\let\svgwidth\undefined%
  \global\let\svgscale\undefined%
  \makeatother%
  \begin{picture}(1,0.15870147)%
    \lineheight{1}%
    \setlength\tabcolsep{0pt}%
    \put(0,0){\includegraphics[width=\unitlength,page=1]{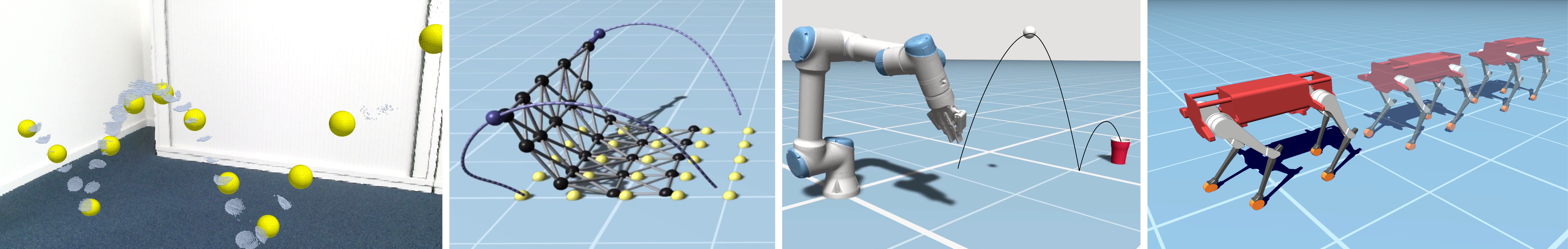}}%
  \end{picture}%
\endgroup%

\caption{
Applications of our differentiable simulation framework (left to right): estimation of stiffness, damping and friction properties from real-world experiments; manipulation of multi-body systems; self-supervised learning of control policies for a throwing task; and physics-based motion planning for robotic creatures with compliant motors and soft feet.
}
\label{fg:teaser}
\end{teaserfigure}

\maketitle


\section{Introduction}

Simulation tools are crucial to a variety of applications in engineering and robotics, where they can be used to test the performance of a design long before the first prototype is ever built.
Based on forward simulation, however, these virtual proving grounds are typically limited to trial-and-error approaches, where the onus is on the user to painstakingly find appropriate control or design parameters.
Inverse simulation tools promise a more direct and powerful approach, as they can anticipate and exploit the way in which a change in parameters affects the performance of the design. 
Unlike forward simulation, this inverse approach hinges on the ability to compute derivatives of simulation runs. 

Differentiable simulations have recently seen increasing attention in the vision, graphics, and robotics communities, e.g., in the context of fluid animation \cite{Holl20Learning}, soft body dynamics \cite{Hu2019}, and light transport \cite{Loubet19Reparametrizing}. 
As a characteristic trait of real-world interactions, however, the strong non-linearities and quasi-discontinuities induced by collisions and frictional contact pose substantial challenges for differentiable simulation. 

The equations of motion for mechanical systems with frictional contact can be cast as a non-linear complementarity problem that couples the tangential forces to the magnitude of the normal forces. While the exact form depends on the friction model being used, this formulation typically postulates different regimes---approaching or separating and stick or slip---with different bounds on the admissible forces. These dichotomies induce discontinuities in forces and their derivatives which, already for the forward problem, necessitate complex numerical solvers that are computationally demanding. For inverse simulations that rely on gradient-based optimization, however, this lack of smoothness is a major stumbling block.

\rev{To overcome the difficulties of the non-smooth problem setting, we propose a differentiable simulator that combines fully implicit time stepping with a principled mollification of normal and tangential contact forces. These contact forces, as well as the coupled system mechanics of rigid and soft objects, are handled through a soft constraint formulation that is simple, numerically robust and very effective. This formulation also allows us to analytically compute derivatives of simulation outcomes through adjoint sensitivity analysis. Our simulation model enables an easy-to-tune trade-off between accuracy and smoothness of objective function landscapes, and we showcase its potential by applying it to a variety of application domains, which can be summarized as follows:
}{
To overcome the difficulties of the non-smooth problem setting, we propose a principled mollification of normal and tangential contact forces with tunable smoothness. 
In combination with implicit integration, our formulation allows for efficient, robust, and fully differentiable time stepping, which we leverage for analytically computing gradients using adjoint sensitivity analysis.
We use these gradients to construct an optimization algorithm that, based on numerical continuation, adapts the smoothness of the contact forces based on progress: initially, a smoother landscape enables rapid progress towards good local minima; gradually increasing sharpness allows for zoning in on solutions with increasing fidelity, approaching the physically-accurate behaviour in the limit.

We validate our contact model on targeted experiments and evaluate the impact of individual design choices by comparing to alternative strategies.
Our formulation for differentiable dynamics is flexible and versatile, which we demonstrate on a diverse set of applications as demonstrated in our results.}

\emph{Real2Sim}: We perform data-driven parameter estimation to find material constants that characterize the stiffness, viscous damping, and friction properties of deformable objects. This enables the creation of simulation models that accurately reproduce and predict the behavior of physical specimens. 

\emph{Control}: Our differentiable simulator is ideally suited for control problems that involve coupled dynamical systems and frictional contact. We explore this important problem domain in the context of robotic manipulation, focusing, through simulation and physical experiments, on behaviors that include controlled multi-bounce tossing, dragging, and sliding of different types of objects. We also use our simulator to generate optimal motion trajectories for a simulated legged robot with compliant actuators and soft feet.
\rev{}{Combining differentiable simulation with continuation on the contact penalty factor allows us to quickly design physically reproducible motions.}

\emph{Self-supervised learning}: By embedding our simulator as a specialized node in a neural network, we show that control policies that map target outcomes to appropriate control inputs can be easily learned in a self-supervised manner; this is achieved by defining the loss directly as a function of the results generated with our differentiable simulation model. We explore this concept through a simple game where a robot arm learns how to toss a ball such that it lands in an interactively placed cup after exactly one bounce.

Our main contributions can be summarized as follows:
\begin{itemize}
\item A stable, differentiable, two-way coupled simulation framework for both rigid and soft bodies.
\item A smooth frictional contact model that can effectively balance differentiability requirements for inverse problems with accurate modelling of real-world behaviour.
\item An evaluation of our friction formulations on a diverse set of applications that highlight the benefits of differentiable simulators.
\end{itemize}

\section{Related work}



\rev{Contact modeling is a well-studied problem in mechanics (see, e.g., ~\cite{bernard1999}), and has received a significant amount of attention in graphics due to its importance in physics-based animation. Here, we focus our review on frictional contact dynamics, its differentability, and its use in solving inverse problems.}{}

\paragraph{Frictional Contact Dynamics} For rigid bodies, early work focused on acceleration-level~\cite{Baraff1994} and velocity-level~\cite{Stewart1996,Anitescu1997} linear complementarity problem (LCP) formulations, and isotropic Coulomb friction. Later, iterative LCP approaches were explored~\cite{Duriez2006,Erleben2007}, and LCP formulations extended to quasi-rigid objects~\cite{Pauly2004}, deformable solids~\cite{Duriez2006,Otaduy2009}, and for fast simulation of large sets of rigid bodies~\cite{Kaufman2005}. Harmon et al.~\shortcite{Harmon2009} propose an asynchronous treatment of contact mechanics for deformable models, discretizing the contact barrier potential. Unifying fast collision detection and contact force computations, Allard et al.~\shortcite{Allard2010} target deformable contact dynamics where deeper penetrations can arise. Exact Coulomb friction paired with adaptive node refinement at contact locations is used for accurate cloth simulation~\cite{Li2018}. Anisotropic Coulomb friction in the deformable~\cite{Pabst2009}, and rigid body setting~\cite{Erleben2019}, has also been considered.

Like Kaufman et al.~\shortcite{Kaufman2008}, our frictional contact formulation applies to coupled rigid and deformable bodies. For this setting, Macklin et al.~\shortcite{Macklin2019} solve a non-linear complementarity problem (NCP) for an exact Coulomb friction formulation with a non-smooth Newton method. A similar technique was applied to fiber assemblies~\cite{BertailsDescoubes2011}. However, in contrast to the above body of work, we target inverse contact applications, and not animation~\cite{Twigg2007,Coros2012,Tan2012}. Interestingly, our hybrid approach can be interpreted as the opposite of staggered projections~\cite{Kaufman2008}, as we soften contact constraints with penalty forces, but can enforce static friction with hard constraints. Softening contact has also proven useful for the control of physics-based characters~\cite{Jain2011}. 

In the context of identifying frictional parameters, we share goals with Pai et al.~\shortcite{Pai2001}. However, in contrast to measuring frictional textures with a robotic system, we aim at characterizing material properties and initial conditions from motion capture (MoCap) data. For parameter estimation from MoCap, we draw inspiration from physics-guided reconstruction~\cite{Monszpart2016} who consider frictionless contact between rigid bodies. Our formulation extends to frictional contact and deformable bodies.

\paragraph{Inverse Contact} Ly et al.~\shortcite{Ly2018} solve for the rest configuration and friction forces of a shell model such that the corresponding static equilibrium state best approximates a user-specified target shape. In a first pass, contact is handled using frictionless bilateral constraints whereas frictional forces are found in a second pass. 
While Ly et al. focus on quasi-static inverse problems, our focus is on inverse \emph{dynamics}. 
To enable design optimization, Chen et al.~\shortcite{Chen2017} introduced an implicitly integrated elastodynamics solver that can handle complex contact scenarios. We share the goal of making our frictional contact model smooth and invertible with previous work in optimal control~\cite{Todorov2011,Erez2012}. However, instead of relying on finite-differencing and explicit time integration~\cite{Todorov2012}, our approach builds on analytically-computed derivatives and fully implicit time integration, allowing us to handle both soft and stiff systems.


\paragraph{Differentiable Simulation} 

An emerging class of differentiable physics-based simulators are increasingly used for applications in machine learning, physics-based control, and inverse design.
While commercial physics engines such as NVIDIA's PhysX or Bullet Physics enjoy widespread use, they prioritize speed over accuracy, especially in the context of dynamics of deformable objects. 
The \hbox{``MuJoCo''} framework \cite{Todorov2012} has been designed as a differentiable simulator for articulated rigid body dynamics. However, derivatives are computed through finite differences, and the use of an explicit time integration method in conjunction with a penalty-based contact model restricts the size of the time steps that can be used for reasonably accurate results. Targeting end-to-end learning, de Avila Belbute-Peres et al.~\shortcite{deAvila2018} show how to compute analytical gradients of LCPs for simple rigid-body systems. In contrast, our approach is able to handle multi-body systems that couple rigid and deformable objects, and it is very robust with respect to the coefficients used for the contact model even under large time steps.
\rev{Degrave et al.~\shortcite{Degrave2019} rely on auto-differentiation to compute derivatives of rigid-body dynamics. However, contact modeling is restricted to spheres and planes.}{} Hu et al.~\shortcite{Hu2019} base their differentiable simulation on the material point method and target specific applications in soft robotics for which explicit time integration is sufficient. In targeting end-to-end learning of controllers, recording and reversing simulations for back-propagation, \rev{DiffTaichi~\shortcite{Hu2019difftaichi} also rely on auto-differentiation. Our approach, on the other hand, analytically computes derivatives via sensitivity analysis.
}{} 
The approach we use is related to recent work in graphics~\cite{Hoshyari2019,Hahn2019} which describes differentiable simulators for viscoelastic materials and constrained flexible multibody dynamics. However, as a significant departure from prior work, frictional contact is at the core of our paper. 
\rev{More precisely, we propose an analytically differentiable simulator that handles frictional contact for rigid and deformable objects within a unified framework. The easy-to-tune characteristics of our model make it applicable to a wide range of inverse dynamics problems.}

\section{Differentiable Multi-Body Dynamics: Preliminaries} \label{sc:theory}

\rev{We begin by deriving the mathematical formulation underlying our analytically differentiable simulation model.
While this derivation follows general concepts recently described in the literature~\cite{Hahn2019,Zimmermann2019}, we note that our quest for a unified simulation framework for multi-body systems composed of arbitrary arrangements of rigid and soft objects demands a somewhat different methodology, as detailed below.}{}

\subsection{Implicit time-stepping for multi-body systems}

We start from a time-continuous dynamics problem described in terms of generalized coordinates ${\mathbf{q}(t)}$, see also \cite{Kaufman2008}.
These generalized coordinates can directly represent the nodal positions of a finite element mesh or the degrees of freedom that describe the position and orientation of a rigid body in space.
\rev{Without loss of generality, here we consider dynamical systems that are characterized by a set of \emph{time independent} input parameters $\mbf{p}$, which could represent, for instance, material constants or a fixed sequence of control actions.}{}
The dynamics of such a system can be succinctly represented through a differential equation of the form:
\[{\mathbf{r}}({\mathbf{q}},\dot {\mathbf{q}},\ddot {\mathbf{q}}, \mbf{p}) = {\mathbf{0}} \quad \forall t, \textrm{where } 
{{\mathbf{r}}}: = {\mathbf{\hat M}}({\mathbf{q}}, \mbf{p})\ddot {\mathbf{q}} - {\mathbf{\hat f}}({\mathbf{q}},\dot {\mathbf{q}}, \mbf{p}) . \eqn \label{eq:rDyn}
\] 

For soft bodies, the generalized mass ${\mathbf{\hat M}}$ is identical to the constant FEM mass matrix, whereas for each rigid body, the corresponding $6\times 6$ block is defined as 
\[ \mbf{\hat M}_\text{RB} := \left( {\begin{array}{*{20}{c}} {m{\mathbf{I}}}&0 \\0&{{\mathbf{I}}_q} \end{array}} \right), \]
with $m$ and ${\mathbf{I}}_q$ representing its mass and configuration-dependent moment of inertia tensor in generalized coordinates~\cite{liu2012quick} (see also our supplements).

The generalized force vector ${\mathbf{\hat f}}$ aggregates all internal and external forces applied to the simulation's degrees of freedom.
Sections~\ref{sc:contact} and \ref{sc:details} detail the exact mathematical formulation for these forces, but here we note that for each rigid body, forces must be mapped from world to generalized space, and an additional term ${\mathbf{C}}({\mathbf{q}},\dot{\mbf{q}})$ that captures the effect of fictitious centrifugal and Coriolis forces must also be added.

Turning to a time-discrete setting, we opt to discretize the residual ${\mathbf{r}}$ using implicit numerical integration schemes.
This decisions warrants a brief discussion.
While implicit integration schemes are standard when it comes to simulation of elastic objects, rigid body dynamics is typically handled with explicit methods such as symplectic Euler.
This misalignment in the choice of integrators often necessitates multi-stage time-stepping schemes for two-way coupling of rigid bodies and elastic objects~\cite{DBLP:conf/sca/ShinarSF08}.
Although effective, such a specialized time-stepping scheme complicates matters when differentiating simulation results.
Furthermore, as discussed in \S~\ref{sc:contact}, the smooth contact model we propose relies on numerically stiff forces arising from a unilateral potential.
Robustly resolving contacts in this setting demands the use of implicit integration for both rigid bodies and elastic objects.
As an added bonus, two-way coupling between these two classes of objects becomes straightforward and numerically robust even under large simulation time steps.

We begin by approximating the first and second time derivatives of the system's generalized coordinates, ${\dot {\mathbf{q}}}$ and ${\ddot {\mathbf{q}}}$, according to an implicit time-discretization scheme of our choice.
For instance with BDF1 (implicit Euler), they are \rev{$\dot {\mathbf{q}}(t_i) \approx \dot{\mbf{q}}^i = ({{\mathbf{q}}^ {i} } - {{\mathbf{q}}^{i-1}})/{\Delta _t}$ and $\ddot {\mathbf{q}}(t_i) \approx \ddot{\mbf{q}}^i = ({{\mathbf{q}}^{i} } - 2{{\mathbf{q}}^{i-1}} + {{\mathbf{q}}^{i-2} })/\Delta _t^2$, where ${\Delta _t}$ is a constant finite time step and superscript $i$ refers to the time step index for time $t_i$.}{}

For each forward simulation step $i$ we then employ Newton's method to find the end-of-time-step configuration ${{\mathbf{q}}^{i} }$ such that ${\mbf{r}^i := {\mathbf{r}}({\mathbf{q}^i},\dot {\mathbf{q}}^i,\ddot {\mathbf{q}}^i, \mbf{p}) = {\mathbf{\hat M}}({\mathbf{q}^{i}}, \mbf{p})\ddot {\mathbf{q}}^{i} - {\mathbf{\hat f}}({\mathbf{q}}^{i},\dot {\mathbf{q}}^{i}, \mbf{p}) = 0}$.
This process requires the Jacobian ${{d{{\mathbf{r}}^i}}/{d{\mathbf{q}}^i}}$, which combines 
derivatives of the generalized forces, the discretized generalized velocities and accelerations, as well as the generalized mass matrix w.r.t.~${{\mathbf{q}}^{i} }$.
Recall that for rigid bodies the mass matrix is configuration-dependent, see also our supplements for details.
We note that in contrast to previous work~\cite{Hahn2019,Zimmermann2019}, we cannot treat time stepping as an energy minimization problem due to this state dependent nature of the mass matrix.
Nevertheless, in conjunction with a line-search routine that monitors the magnitude of the residual, directly finding the root of $\mbf{r}^i$ works very well in practice.

\subsection{Simulation Derivatives}

\sloppy
With the basic procedure for implicit time-stepping in place, we now turn our attention to the task of computing derivatives of simulation outcomes.
To keep the exposition brief, we concatenate the generalized coordinates for an entire sequence of time steps computed through forward simulation into a vector ${{\mathbf{\tilde q}}: =  {({{\mathbf{q}}^{1\user1{T}}}, \ldots ,{{\mathbf{q}}^{{n_t}}}^\user1{T})^\user1{T}} }$, where $n_t$ is the number of time steps.
Similarly, we let ${\mathbf{\tilde r}}$ be the vector that concatenates the residuals for all time steps of a simulation run.
\rev{
Differentiating ${\mathbf{\tilde r}}$ with respect to input parameters $\mbf{p}$ gives:
\[
\frac{{d{{\mathbf{\tilde r}}}}}{{d{\mathbf{p}}}} = \frac{{\partial {{\mathbf{\tilde r}}}}}{{\partial {\mathbf{p}}}} + \frac{{\partial {{\mathbf{\tilde r}}}}}{{\partial {\mathbf{\tilde q}}}} \frac{{d {{\mathbf{\tilde q}}}}}{{d{\mathbf{p}}}}. \eqn \label{eq:drdpTotal}
\]
Since for \emph{any} $\mbf{p}$ we compute a motion trajectory such that ${\mathbf{\tilde r}} = \mathbf{0}$ during the forward simulation stage, the total derivative \eqref{eq:drdpTotal} is also $\mathbf{0}$.
This simple observation lets us compute the Jacobian ${{d {{\mathbf{\tilde q}}}}}/{{d{\mathbf{p}}}}$, which is also called the sensitivity matrix, as:
\[\mathbf{S} := \frac{{d {{\mathbf{\tilde q}}}}}{{d{\mathbf{p}}}} = - \left({\frac{{\partial {{\mathbf{\tilde r}}}}}{{\partial {\mathbf{\tilde q}}}}}\right) ^{-1}
\frac{{\partial {{\mathbf{\tilde r}}}}}{{\partial {\mathbf{p}}}}. \eqn
\]
The sensitivity matrix $\mathbf{S}$ captures the way in which an entire trajectory computed through forward simulation changes as the input parameters $\mbf{p}$ change.
Note that $\mbf{S}$ can be computed efficiently by exploiting the sparsity structure of ${{\partial {{\mathbf{\tilde r}}}}}/{{\partial {\mathbf{\tilde q}}}}$, a matrix that can be easily evaluated by re-using most of the same ingredients already computed for forward simulation.
For example, in the case of BDF1, each sub-block $\mathbf{s}^i_j := {{{d {{\mathbf{q}}^i}}}/{{d {\mathbf{p}}_j}}}$ of $\mathbf{S}$, which represents the change in the configuration of the dynamical system at time $t_i$ with respect to the $j$-th input parameter $\mbf{p}_j$, reduces to:
\[
{\mathbf{s}}^i_j = - \left({\frac{{\partial {{\mathbf{ r}}^i}}}{{\partial {\mathbf{q}}^i}}}\right) ^{-1} 
\left( {\frac{{\partial {{\mathbf{ r}}^i}}}{{\partial {\mathbf{p}}_j}}} + 
{\frac{{\partial {{\mathbf{r}}^i}}}{{\partial {\mathbf{q}}^{j-1}}}}
\mathbf{s}^{i-1}_j + 
{\frac{{\partial {{\mathbf{r}}^i}}}{{\partial {\mathbf{q}}^{j-2}}}}
\mathbf{s}^{i-2}_j
\right). \eqn
\]
Note that the dependency on past sensitivities follows the same time stepping scheme as the forward simulation.
We refer the interested reader to~\cite{Simonv2} for more details on this topic, but for completeness we briefly describe a variant of this formulation, the adjoint method, as it can lead to additional gains in computational efficiency.
}{}

\subsection{The adjoint method}
For most applications we do not need to compute the sensitivity matrix explicitly.
Instead what we are interested in, is finding input parameters $\mathbf{p}^*$ that minimize an objective defined as a function of the simulation result:
\[{{\mathbf{p}}^*} = \arg {\min _{\mathbf{p}}} \; \Phi \left(\, {\mathbf{p}},{\mathbf{\tilde q}(\mathbf{p})} \, \right), \eqn \label{eq:phi} \]
where ${\mathbf{\tilde q}(\mathbf{p})}$ is the entire trajectory of the dynamical system.

In order to solve this optimization problem using gradient based methods such as ADAM \cite{Kingma2014,Optimlib2019} or L-BFGS \cite{Nocedal1980,Lbfgspp2019}, we need to compute the derivative 
\[\frac{{d\Phi }}{{d{\mathbf{p}}}} = \frac{{\partial \Phi }}{{\partial {\mathbf{p}}}} + \frac{{\partial \Phi }}{{\partial {\mathbf{\tilde q}}}} {{\mathbf{S}}}.
\eqn \label{eq:dPhi} \]

The adjoint method enables the computation of this objective function gradient without computing $\mbf{S}$ directly.
\rev{While Bradley~\shortcite{Bradley2013} derives the adjoint method for the time-continuous case, here we describe it directly in the time-discretized setting.}{}

\sloppy
We first express the sensitivity matrix as ${{\mathbf{S}}} =  - {{\mathbf{\tilde B}}^{ - 1}}{\mathbf{\tilde A}}$\rev{, with ${\mathbf{\tilde A} := {{\partial {{\mathbf{\tilde r}}}}}/{{\partial {\mathbf{p}}}}}$ and ${\mathbf{\tilde B} := {{\partial {{\mathbf{\tilde r}}}}}/{{\partial {\mathbf{\tilde q}}}}}$, respectively.
Introducing the notation ${\mbf{\tilde y} := {\partial \Phi / \partial \mathbf{\tilde q}}}$, Eq.~\eqref{eq:dPhi} becomes}{}
\[\frac{{d\Phi }}{{d{\mathbf{p}}}} = \frac{{\partial \Phi }}{{\partial {\mathbf{p}}}} - {\mathbf{\tilde y}}{{\mathbf{\tilde B}}^{ - 1}}{\mathbf{\tilde A}}, \eqn \label{eq:directGrad}\]
Instead of evaluating this expression directly, we first solve for the \emph{adjoint state} $\mbf{\tilde \lambda}$:
\[{{\mathbf{\tilde B}}^\user1{T}}{\mbf{\tilde \lambda }} = {{\mbf{\tilde y}}^\user1{T}}, \eqn \label{eq:adjoint} \]
and then compute the objective function gradient as:
\[\frac{{d\Phi }}{{d{\mathbf{p}}}} = \frac{{\partial \Phi }}{{\partial {\mathbf{p}}}} - {{\mbf{\tilde \lambda }}^\user1{T}}{\mathbf{\tilde A}}. \eqn \label{eq:adjGrad} \]

Note that both ${\mbf{\tilde y}}$ and ${\mbf{\tilde \lambda}}$ are vectors of length $\left| {\mathbf{q}} \right|{n_t}$, whereas the size of ${\mbf{\tilde A}}$ is $\left| {\mathbf{q}} \right|{n_t} \times \left| {\mathbf{p}} \right|$, where $n_t$ is the number of time steps, $\left| {\mathbf{q}} \right|$ is the number of degrees of freedom, and $\left| {\mathbf{p}} \right|$ is the number of parameters.
Furthermore, the sparsity of ${\mathbf{\tilde A}}$ (or lack thereof) depends on the time interval and spatial region where each parameter influences the simulation.
For instance, homogeneous material parameters of deformable objects generally affect the entire simulation and lead to dense (parts of) ${\mathbf{\tilde A}}$ , whereas control inputs for specific time steps influence only a small subset of the simulation and result in a much sparser structure. As a final note, it is worth mentioning that the block-triangular structure of ${\mathbf{\tilde B}}$ can also be leveraged to speed up the computation of the adjoint state.
\rev{While this block-matrix form emphasizes when (and where) data must be stored during the forward simulation, regardless of how the system is solved, while computing the adjoint state, information propagates backwards in time (a common feature of time-dependent adjoint formulations).}{}

\section{Differentiable Frictional Contact Model} \label{sc:contact}


With the differentiable simulation framework in place, our challenge now is to  formulate a smooth frictional contact model that approaches, in the limit, the discontinuous nature of physical contacts.
\rev{In this context, we always refer to the smoothness of the resulting trajectory, rather than the contact forces.
Specifically, an impulsive force would result in a non-smooth trajectory, which we must avoid.}{}
As we show in this section, a soft constraint approach to modelling frictional contact fits seamlessly into the theory presented in \S~\ref{sc:theory}, and enables an easy-to-tune trade-off between accuracy (e.g. solutions satisfying contact complementarity constraints and Coulomb's law of friction) and smoother, easier to optimize objective function landscapes.

We begin by formulating the contact conditions and response forces in terms of \rev{a single contact point}{the contact location} $\mbf{x}(\mbf{q})$. 
For deformable objects, we handle contacts on the nodes of the FEM mesh.
For rigid bodies, we implement spherical or point-based collision proxies and then map the resulting world-space contact forces \rev{$\mbf{f}$}{} into generalized coordinate space \rev{$\mbf{\hat f}$}{}, as described in \S~\ref{sc:details}.

For the normal component of the contact we must find a state of the dynamical system such that
\[g(\mathbf{x}) \geq 0, \eqn \label{eq:gap} \]
where we refer to $g$ as the \emph{gap} function, measuring the distance from $\mathbf{x}$ to the closest obstacle.
Assuming that $g$ is a signed distance function (and negative if $\mathbf{x}$ lies inside of an obstacle), we define the outward unit normal ${\mathbf{n}}: = (\partial g/\partial {\mathbf{x}}) ^\user1{T}$.
\rev{Note that we use the convention that the derivative of a scalar function w.r.t.~a vector argument is a row-vector throughout this paper, consequently $\mbf{n}$ is a column-vector.
Furthermore, we assume that $g$ is available in closed form; we mostly use planar obstacles in our examples.}{}
We will also use the notation ${\mathbf{N}}: = {\mathbf{n}}{{\mathbf{n}}^\user1{T}}$ for the matrix projecting to this normal.
The force required to maintain non-negative gaps (non-penetration), denoted ${{\mathbf{f}}_n}$, must be oriented along ${\mathbf{n}}$, i.e.~${{\mathbf{f}}_n} = {f_n}{\mathbf{n}}$.
The \emph{normal force magnitude} ${f_n}$ must be non-negative and can only be non-zero if there is a contact, consequently (Hertz-Signorini-Moreau condition, see also Eq.~(2.10) in \cite{Wriggers2006}):
\[{f_n} \geq 0, \; f_n \, g = 0 \eqn \label{eq:signorini}\]

In the tangential direction, we must determine a friction force ${{\mathbf{f}}_t}$ whose magnitude is constrained by the Coulomb limit
\[\left\| {{{\mathbf{f}}_t}} \right\| \leq {c_f}{f_n}, \eqn \label{eq:coulomb} \]
where $c_f$ is the \emph{coefficient of friction}.
This inequality distinguishes two regimes: \emph{static} and \emph{dynamic} friction (also referred to as \emph{stick} and \emph{slip} respectively).
In the former case, the friction force magnitude is below the Coulomb limit and the tangential velocity vanishes (stick).
In the latter case,\rev{ following the principle of maximum dissipation \cite{Stewart00},}{} the friction force is oriented opposite the tangential velocity and Eq.~\eqref{eq:coulomb} is satisfied as an equality (slip).

More formally we have either
\[{\mathbf{T}\dot {\mathbf{x}}} = 0 \text{ (stick), or} \eqn \label{eq:stick} \]
\[{{\mathbf{f}}_t} =  - {c_f}{f_n}{\mathbf{t}} \text{ (slip),} \eqn \label{eq:slip} \]
where $\mathbf{t}:=({\mathbf{T}\dot {\mathbf{x}}})/\left\| {{\mathbf{T}\dot {\mathbf{x}}}} \right\|$ and ${\mathbf{T}}: = {\mathbf{I}} - {\mathbf{N}}$ projects to the tangent plane.
Note that in either case ${{\mathbf{f}}_t}\cdot{\mathbf{n}} = 0$ must hold.

\subsection{Sequential Quadratic Programming} \label{sc:sqp}

\rev{Here, we briefly outline a hard constrained formulation of frictional contacts, which we employ for comparison.}{}
Introducing Lagrange multipliers (representing contact forces) for the contact constraints \eqref{eq:gap} and \eqref{eq:stick}, along with the dynamic friction forces, into the residual \eqref{eq:rDyn} leads to a system of equations representing the Karush-Kuhn-Tucker conditions of an optimization problem:
\[\begin{gathered}
  {\mathbf{r}} + {{\mathbf{n}}}{{\lambda }} + {{\mathbf{\bar T}}^\user1{T}}{\mbf{\mu }} = 0, \\
  g{\text{(}}{\mathbf{x}}{\text{)}} \geq 0,{\text{ and}} \\
  {\mathbf{\bar T}\dot{\mbf{ x}}} = 0, \end{gathered} \eqn \label{eq:KKT} \]
where ${\lambda}$ and ${\mbf{\mu }}$ are the Langrange multipliers for the normal and static friction constraints respectively\rev{.
For a single contact point in static friction, ${\mbf{\mu }}$ has two components and $\mbf{\bar T}$ contains two orthogonal tangent vectors (corresponding to the two non-zero eigenvalues of ${\mathbf{T}}$).
In the case of dynamic friction (sliding) we remove the constraint on the tangential velocity and replace the friction force ${{\mathbf{\bar T}}^\user1{T}}{\mbf{\mu }}$ with $\mbf{f}_t$ according to Eq.~\eqref{eq:slip}.
}{ 
, and consequently 
${\mathbf{n}}{f_n} = {{\mathbf{N}}^\user1{T}}{\mbf{\lambda }}$, and similarly 
${\mathbf{f}}_t = {{\mathbf{T}}^\user1{T}}{\mbf{\mu }}$ for the tangential sticking forces.}

In principle, the KKT system \eqref{eq:KKT} can be linearized and solved as a sequence of quadratic programs (SQP).
\rev{}{Each iteration improves the current estimate, around which the next linearization is constructed.}
In this way, each quadratic program takes the role of the linear solve in Newton's method.
However, there are some limitations to this approach.
In particular, linearizing the dynamic friction force resulting from Eq.~\eqref{eq:slip} is not straightforward, as this force depends not only on the simulation state and velocity, but also\rev{ on the normal force, in this case given by the Lagrange multiplier $\lambda$.}{ on the normal force $f_n$, which results from the Lagrange multipliers $\mbf{\lambda}$.}
Furthermore, the direction of the friction force should be opposing the current sliding velocity, which is ill-defined if this velocity approaches zero.

One way to address these issues is to assume that $\mbf{f}_t$ is constant rather than linear, and compute its value based on the previous iteration.
Doing so however reduces the convergence of the SQP iteration.
In some cases the ill-conditioned behaviour of the unit vector $\mbf{t}$ can even cause this approach to fail to converge at all.
To help mitigate this problem, we first approximately solve the problem assuming sticking contacs, and then only update the friction force magnitude according to the Coulomb limit, but maintain its direction.
\rev{Similarly, the approach of Tan et al.~\shortcite{Tan2012} formulates the frictional contact problem as a quadratic program with linear complementarity constraints by linearizing the friction cone. Their solver also iteratively updates a set of active constraints, which effectively selects the direction of the sliding force.}{}
Finally, the SQP iteration is not as straightforward to stabilize by a line-search procedure as a Newton iteration would be.
Macklin et al.~\shortcite{Macklin2019} report similar issues and present various approaches to improve upon them.

Note that derivatives of the KKT conditions \eqref{eq:KKT}, once the solver has converged, could still be used to compute simulation state derivatives by direct differentiation.
However, it is currently not clear whether this approach would immediately integrate with our \emph{adjoint} formulation.
We refer to \cite{Amos2017} for further details on derivatives of QP solutions.

In the light of these limitations, we have implemented an SQP forward simulation approach only for the sake of comparisons on simple deformable examples, where stability and convergence is not an issue; see also Fig.~\ref{fg:CylinderDrop}. We show that our hybrid method, \S\ref{sc:hybridContacts}, yields visually indistinguishable results.

\subsection{Penalty methods}

Another approach to solve dynamics with contacts is to convert the constraints \eqref{eq:gap} and \eqref{eq:stick} to \emph{soft} constraints, and introduce penalty forces if the constraints are violated.
For the contact constraint, this means we need to add a force that is oriented along the outward normal and vanishes when $g>0$.
There are various types of functions, such as soft-max or truncated $\log$-barriers, that we can use for this purpose.
We find that even the simplest choice of a piece-wise \emph{linear penalty function} is fast and sufficiently accurate for our applications:
\[{{\mathbf{f}}_n} = {\mathbf{n}}\,{k_n}\max (-g, \; 0), \eqn \label{eq:linPenN} \]
where $k_n$ is a penalty factor that must be chosen large enough to sufficiently enforce the normal constraint.

\rev{
Note that $k_n$ effectively controls the steepness of the resulting force and therefore the smoothness of the collision response.
Of course the derivative of this force contains a discontinuity due to the $\max$ operator. While this could easily be replaced by a soft-max, our experiments indicate that doing so is not necessary. To find itself exactly at the kink in the contact force, a point on the multi-body system would have to be on the $g=0$ manifold at the end of the time step due only to gravity, inertia, and internal force, which is extremely unlikely. The force Jacobian, which is needed for both forward simulation and to compute derivatives is well-defined everywhere else.
}{}

Similarly, we can formulate Coulomb friction as a clamped linear \emph{tangential} penalty force with corresponding penalty factor $k_t$:
\[{{\mathbf{f}}_t} =  - {\mathbf{t}}\min ({k_t}\left\| {{\mathbf{T}\dot{\mathbf{ x}}}} \right\|,\;{c_f}{f_n}). \eqn \label{eq:linPenT} \]
This expression reduces to \eqref{eq:slip} for dynamic friction, but results in the linear penalty force ${k_t \mathbf{T}\dot{\mathbf{ x}}}$ for static friction.

Instead of solving the KKT conditions \eqref{eq:KKT}, we now only need to solve a formally unconstrained system that has the same form as Eq.~\eqref{eq:rDyn}, with additional penalty forces according to \eqref{eq:linPenN} and \eqref{eq:linPenT}:
${\mathbf{r}} + {{\mathbf{f}}_t} +{{\mathbf{f}}_n} =0.$
From now on, we assume that these penalty forces (mapped to generalized coordinates in the case of rigid bodies) are part of the residual itself.
Consequently, we can still use a standard Newton method to solve for the dynamic behaviour (after applying an adequate time discretization scheme).
Stabilizing the solver with a line-search method that enforces decreasing residuals is sufficient to handle the non-smooth points of the penalty forces introduced by the $\max$ and $\min$ operators.

We can also smooth out the transition between stick penalty and dynamic friction force and replace $\mbf{f}_t$ of Eq.~\eqref{eq:linPenT} with
\[ {{\mathbf{f}}_t} =  - {\mathbf{t}}\,{c_f}{f_n}\tanh \left(\, {{k_t}} \left\| {{\mbf{T}\dot{\mbf{x}}}} \right\| /{({c_f}{f_n})} \,\right). \eqn \label{eq:tanhPenT} \]
Using a hyperbolic tangent function introduces slightly softer friction constraints and can help improve the performance of optimization methods.
\rev{Note that for tangential velocities close to zero $\tanh (v) \approx v$, which resolves any numerical issues with computing $\mbf{t}$ for small velocities.}{}
\rev{In our examples we always choose $k_t=k_n \Delta_t$.}{}

\subsection{Hybrid method} \label{sc:hybridContacts}

\rev{

Solving frictional contacts with hard constraints is both computationally expensive in terms of the forward simulation, as well as more challenging to differentiate, especially in an adjoint formulation.
Typically, the normal contact constraint is satisfied exactly, but the friction model is simplified to enable more efficient simulation.
Even then, the resulting complementarities are technically a combinatorial problem and heuristics are employed to choose the active constraints.

Penalty formulations, on the other hand, are formally unconstrained and therefore relatively straightforward to simulate and differentiate.
While the choice of penalty stiffness does affect numerical conditioning, using implicit integration enables stable simulation for a wide range of stiffness.
The main drawback of penalties is that one must allow some small constraint violations.

While a small violation of the normal constraint \eqref{eq:gap} is often visually imperceptible, 
softening the static friction constraint can introduce unacceptable artefacts in some situations.
In particular, if a (heavy) object is supposed to \emph{rest} on an inclined surface \emph{under static friction} and the simulation runs for a sufficiently long time, the tangential slipping introduced by softening the stick constraint will inevitably become visually noticeable.

We address this problem by formulating a hybrid method using linear penalty forces, Eq.~\eqref{eq:linPenN} and \eqref{eq:linPenT}, combined with \emph{equality} constraints for the static friction case.
In order to take one time step in the forward simulation, we proceed as follows:
first, we approximately solve the linear penalty problem to a residual of ${\left\| {\mathbf{r}} \right\| < {\varepsilon ^{1/2}}}$.
Then, we apply hard constraints of the form \eqref{eq:stick} if the friction force according to \eqref{eq:linPenT} is below the Coulomb limit.
Note that these \emph{equality} constraints are analogous to standard Dirichlet boundary conditions.
Consequently, we continue with the Newton iteration, including these constraints.
However, if enforcing a hard constraint leads to a tangential force exceeding the Coulomb limit, we revert that contact point back to the clamped penalty force \eqref{eq:linPenT}.
We only allow this change if the intermediate state satisfies ${\left\| {\mathbf{r}} \right\| < {\varepsilon ^{1/2}}}$.
In rare cases, we eventually fall back to the penalty formulation.
These cases typically arise right at the transition when a sliding object comes to rest; once at rest the hard constraints keep it in place reliably.
Even if we revert to penalties, we know that hard constraints would have violated the Coulomb limit in that time step, so allowing a small sliding velocity is acceptable at this point.
Finally, we continue the solver iterations until convergence, i.e.~${\left\| {\mathbf{r}} \right\| < {\varepsilon}}$.
\begin{wrapfigure}{r}{4cm}
\hspace{-0.6cm}
\def\svgwidth{4.2cm}
\begingroup%
  \makeatletter%
  \providecommand\color[2][]{%
    \errmessage{(Inkscape) Color is used for the text in Inkscape, but the package 'color.sty' is not loaded}%
    \renewcommand\color[2][]{}%
  }%
  \providecommand\transparent[1]{%
    \errmessage{(Inkscape) Transparency is used (non-zero) for the text in Inkscape, but the package 'transparent.sty' is not loaded}%
    \renewcommand\transparent[1]{}%
  }%
  \providecommand\rotatebox[2]{#2}%
  \newcommand*\fsize{\dimexpr\f@size pt\relax}%
  \newcommand*\lineheight[1]{\fontsize{\fsize}{#1\fsize}\selectfont}%
  \ifx\svgwidth\undefined%
    \setlength{\unitlength}{296.97974396bp}%
    \ifx\svgscale\undefined%
      \relax%
    \else%
      \setlength{\unitlength}{\unitlength * \real{\svgscale}}%
    \fi%
  \else%
    \setlength{\unitlength}{\svgwidth}%
  \fi%
  \global\let\svgwidth\undefined%
  \global\let\svgscale\undefined%
  \makeatother%
  \begin{picture}(1,0.58413946)%
    \lineheight{1}%
    \setlength\tabcolsep{0pt}%
    \put(0,0){\includegraphics[width=\unitlength,page=1]{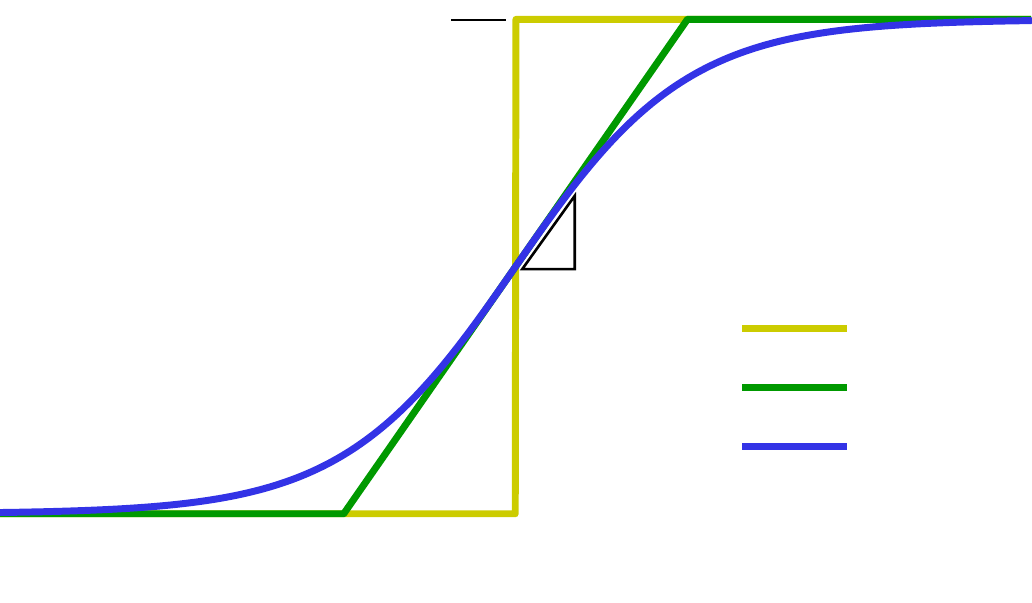}}%
    \put(0.44813175,0.01282612){\color[rgb]{0.14901961,0.14901961,0.14901961}\makebox(0,0)[lt]{\lineheight{1.25}\smash{$ \mbf{T}{\dot{\mbf{x}}} $ }}}%
    \put(0.57998609,0.34852365){\color[rgb]{0.14901961,0.14901961,0.14901961}\makebox(0,0)[lt]{\lineheight{1.25}\smash{$k_t$}}}%
    \put(0.34547436,0.54287535){\color[rgb]{0.14901961,0.14901961,0.14901961}\makebox(0,0)[ct]{\lineheight{1.25}\smash{$c_f \, f_n$}}}%
    \put(0.82853418,0.24761822){\color[rgb]{0,0,0}\makebox(0,0)[lt]{\lineheight{1.25}\smash{\begin{tabular}[t]{l}hybrid\end{tabular}}}}%
    \put(0.82853418,0.19048924){\color[rgb]{0,0,0}\makebox(0,0)[lt]{\lineheight{1.25}\smash{\begin{tabular}[t]{l}linear\end{tabular}}}}%
    \put(0.82853418,0.13336022){\color[rgb]{0,0,0}\makebox(0,0)[lt]{\lineheight{1.25}\smash{\begin{tabular}[t]{l}tanh\end{tabular}}}}%
  \end{picture}%
\endgroup%

\vspace{-\baselineskip}
\end{wrapfigure}
In this way, we enforce static friction with hard constraints whenever possible, while guarenteeing that the Coulomb limit is never violated.

Note that employing a penalty formulation for the \emph{normal} forces means that both, the normal force and the Coulomb limit, depend \emph{directly} on the simulation state (as opposed to including Lagrange multipliers representing normal forces).
Consequently, the derivatives of these forces are straightforward to compute, the solver maintains the quadratic convergence of Newton's method, and can be stabilized with a standard line-search procedure.

Similarly, when computing sensitivities or adjoint objective function gradients, we can incorporate the additional equality constraints in the same way as Dirichlet boundary conditions.
That is, the sensitivities or the adjoint state must fulfil analogous boundary conditions to the forward simulation.
Specifically, for BDF1, the static friction constraint at time $t_i$ becomes ${\mathbf{\bar T}}^i{{\mathbf{x}}^i} = {\mathbf{\bar T}}^i{{\mathbf{x}}^{i - 1}}$ and consequently, we have ${\mathbf{\bar T}}^i \mbf{s}_x^i = {\mathbf{\bar T}}^i \mbf{s}_x^{i - 1}$ for sensitivities of $\mbf{x}$, or ${{\mathbf{\bar T}}^i}{{\mbf{\lambda }}^{i - 1}} = {{\mathbf{\bar T}}^i}{{\mbf{\lambda }}^i}$ for the adjoint state (note that the earlier state is unknown in the latter case).

}{ 
Apart from having to choose a reasonable penalty factor ($f_p$), the main drawback of penalty methods is that they allow some small constraint violations.
While such a small violation of the normal constraint \eqref{eq:gap} is often visually imperceptible (or can be mitigated by introducing a small constant offset to $g$ if necessary), softening the static friction constraint can introduce unacceptable artefacts in some situations.
In particular, if a (heavy) object is supposed to rest on an inclined surface under static friction and the simulation runs for a sufficiently long time, the tangential slipping introduced by softening the stick constraint \eqref{eq:stick} will inevitably become visually noticeable.

One way to address this problem is to apply a penalty method for the normal constraint, i.e.~define the normal force via Eq.~\eqref{eq:linPenN}, but enforce the stick constraint \eqref{eq:stick} exactly. In the dynamic friction regime, the tangential force follows Eq.~\eqref{eq:slip}.
In this way, we ensure physically correct friction behaviour, while simplifying the numerics compared to the SQP approach discussed in \S\ref{sc:sqp}.
In particular, employing a penalty formulation for the normal forces means that both normal force and Coulomb limit only depend on the simulation state directly.
Consequently, the derivatives of these forces are easy to compute, and the solver maintains the quadratic convergence of Newton's method in most cases.
Furthermore, we only add \emph{equality} constraints for static friction, rather than inequalities, which are treated analogous to standard Dirichlet boundary conditions.

However, the distinction between static and dynamic friction now must be made separately in order to apply the correct equality constraints for the sticking contacts.
Within each time step, we employ an \emph{active set} type approach, where we initially assume all nodes that are currently in contact ($g<0$) to stick in the first solver iteration.
If the resulting tangential force exceeds the Coulomb limit for any given node, we label this node as \emph{sliding}, release the equality constraint, and apply the tangential force ${\mbf{f}_t = -\mbf{t} c_f \, f_n }$ instead.
Conversely, if this Coulomb friction force ever results in a change of sliding velocity direction such that $\mbf{t} \cdot \dot{\mbf{x}}^{+} >0$ (at the end of any Newton iteration), we revert the node label to \emph{sticking}.
We allow at most $5$ label changes per node per time step, and the solver terminates only if no node labels were changed in the last iteration (and the residual is sufficiently small).
Note that we need to be careful of numerical issues when computing the tangential direction $\mbf{t}$, as nodes that have had a constraint applied in the previous iteration will have zero (up to machine precision) tangential velocity at the start of the next iteration.
In this case we initialize $\mbf{t}$ to the direction of the previous tangential force (rather than velocity).
We also add a small regularization in subsequent iterations to ensure stable derivatives of the tangential force, i.e.~${\mathbf{t}} = {\mathbf{T}\dot{\mbf{x}}}/(\left\| {{\mathbf{T}\dot{\mbf{x}}}} \right\| + \varepsilon )$, where we use $\varepsilon = 10^{-8}$ in our results.
\begin{wrapfigure}{r}{4cm}
\hspace{-0.6cm}
\def\svgwidth{4.2cm}

\vspace{-\baselineskip}
\end{wrapfigure}
This hybrid method converges faster than the penalty variants for static friction scenarios as the stiff tangential penalty forces are replaced by simple equality constraints (i.e.~additional Dirichlet boundary conditions).
It also converges faster than the SQP approach as the derivatives of the residual can be computed without the need for simplifying assumptions on the coupling between normal and tangential friction forces. 
}

\rev{
\subsection{Summary and evaluation}

\begin{figure}
\def\svgwidth{\columnwidth}
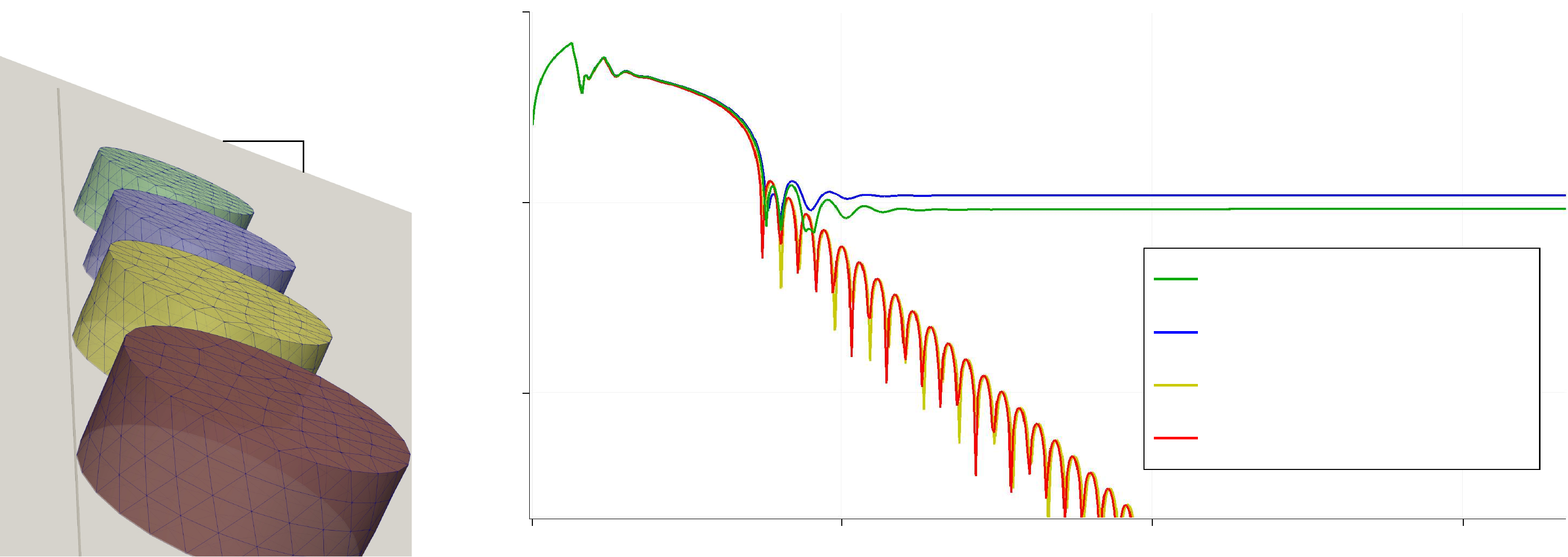
\\ \vspace{0.1cm}
\def\svgwidth{\columnwidth}
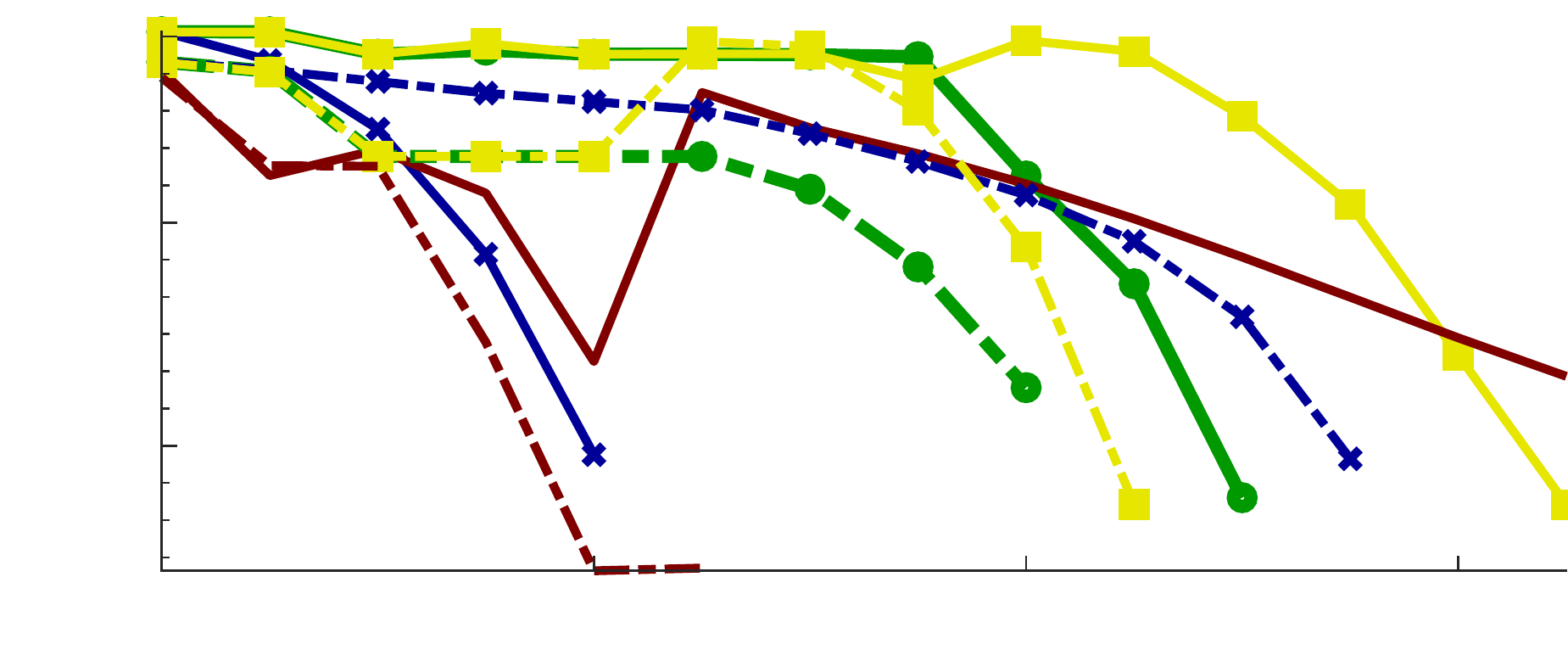 \vspace{-0.8cm}
\caption{Dropping a cylinder onto an inclined plane. The resulting motion is generated using different contact and friction methods: linear penalty (green), tanh penalty (blue), hybrid (yellow), SQP (red).
The final state of the simulation (a) after $500$ time steps ($t=2.5$~s) is shown on the top left, and the mean tangential velocity over time (b) is shown in the top right sub-figure.
Solver convergence for time steps $60$ (sliding phase, solid), and $150$ (sticking phase, dashed) are shown in sub-figure (c).
\rev{}{Note that tangential penalty methods can not achieve a full stop under static friction. Our hybrid method is very close to the SQP solution.}}
\label{fg:CylinderDrop}
\end{figure}

\begin{figure*}
\def\svgwidth{ \textwidth }
\begingroup%
  \makeatletter%
  \providecommand\color[2][]{%
    \errmessage{(Inkscape) Color is used for the text in Inkscape, but the package 'color.sty' is not loaded}%
    \renewcommand\color[2][]{}%
  }%
  \providecommand\transparent[1]{%
    \errmessage{(Inkscape) Transparency is used (non-zero) for the text in Inkscape, but the package 'transparent.sty' is not loaded}%
    \renewcommand\transparent[1]{}%
  }%
  \providecommand\rotatebox[2]{#2}%
  \newcommand*\fsize{\dimexpr\f@size pt\relax}%
  \newcommand*\lineheight[1]{\fontsize{\fsize}{#1\fsize}\selectfont}%
  \ifx\svgwidth\undefined%
    \setlength{\unitlength}{4273.99987793bp}%
    \ifx\svgscale\undefined%
      \relax%
    \else%
      \setlength{\unitlength}{\unitlength * \real{\svgscale}}%
    \fi%
  \else%
    \setlength{\unitlength}{\svgwidth}%
  \fi%
  \global\let\svgwidth\undefined%
  \global\let\svgscale\undefined%
  \makeatother%
  \begin{picture}(1,0.21045864)%
    \lineheight{1}%
    \setlength\tabcolsep{0pt}%
    \put(0,0){\includegraphics[width=\unitlength,page=1]{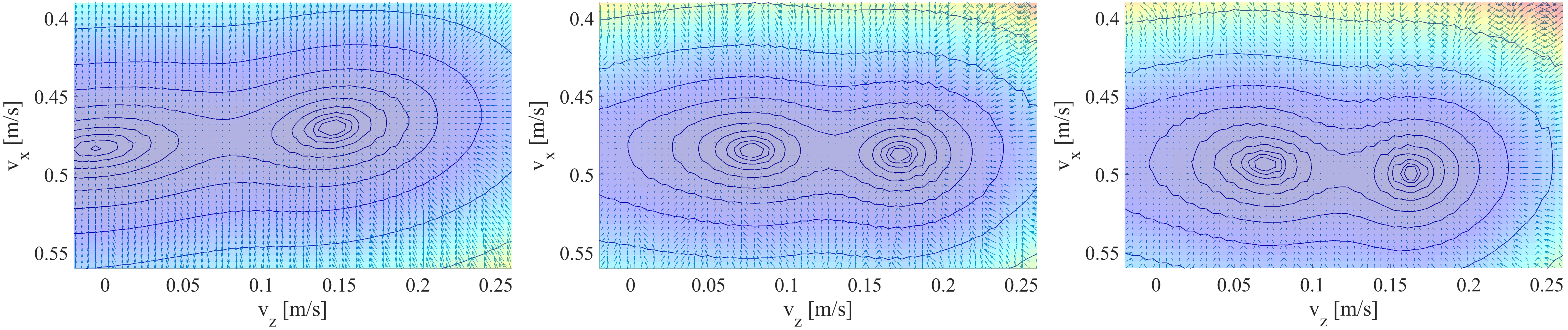}}%
    \put(0.03377252,0.0079249){\color[rgb]{0,0,0}\makebox(0,0)[lt]{\lineheight{1.25}\smash{\begin{tabular}[t]{l}(a)\end{tabular}}}}%
    \put(0.36367429,0.0079249){\color[rgb]{0,0,0}\makebox(0,0)[lt]{\lineheight{1.25}\smash{\begin{tabular}[t]{l}(b)\end{tabular}}}}%
    \put(0.70761436,0.0079249){\color[rgb]{0,0,0}\makebox(0,0)[lt]{\lineheight{1.25}\smash{\begin{tabular}[t]{l}(c)\end{tabular}}}}%
  \end{picture}%
\endgroup%

\caption{
Objective function (shading and isolines) and gradients (arrows) corresponding to the control problem presented in Fig.~\ref{fg:SphereToPointAndLine}a, 
\rev{using soft (a; $k_n=100$) and stiff (b; $k_n=10^3$) parameters for penalty-based contacts, compared to hybrid contacts (c).
Note that the overall nature of the objective landscape (i.e.~the local minima it exhibits) remains the same, although these minima do shift slightly when softening the contacts. This behaviour promotes the use of continuation methods, whereby solutions obtained with a soft, smoother contact model are used as initial guess for simulations with stiffer, more realistic, parameters.
}{with hybrid contacts (a), with tanh penalty contacts and half the time step size (b). Images (c) and (d) show results for a rigid body sphere (instead of deformable FEM) with linear penalty contacts. The time step in (c) is the same as (b), while (d) cuts the time step in half again to $1/4$ of (a) and reduces the penalty factor from $k_n=10^3$ to $k_n=400$. Note that penalty stiffness and time step have a strong impact on the smoothness of the result. Please also refer to our supplemental document for a more detailed comparison.}
}
\label{fg:SphereToPoint-objFcnCombined}
\end{figure*}
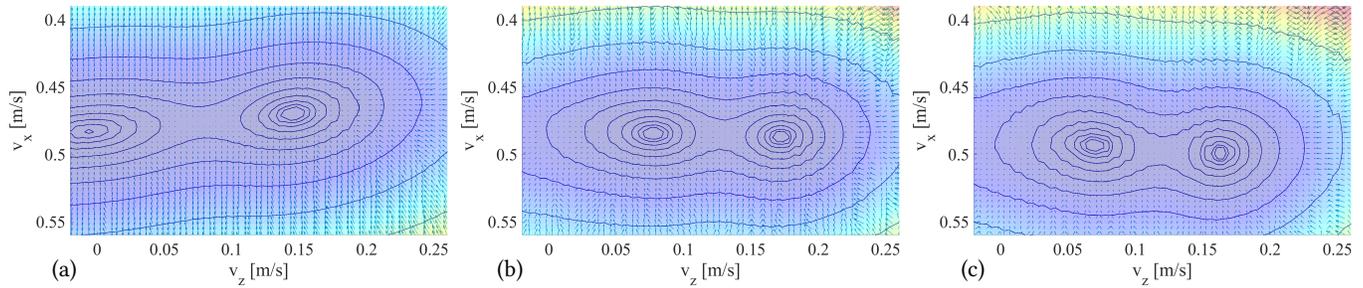

\begin{figure}
\def\svgwidth{\columnwidth}
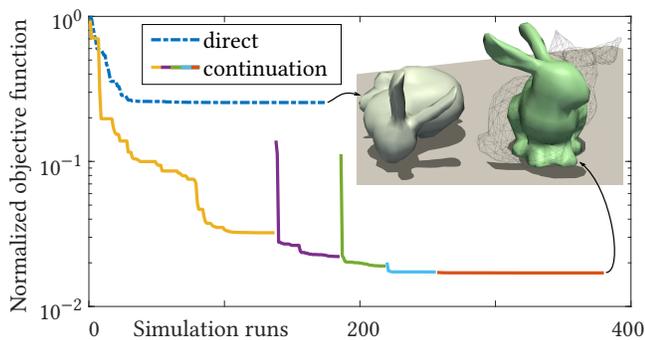
\caption{\rev{
Optimization convergence for throwing a bunny, similar to Fig.~\ref{fg:SphereAndBunnyThrow}b but without the wall, such that it lands upright, close to a target pose. A continuation strategy applied to the stiffness of the contact model can drastically improve optimization results for such challenging control problems.
Direct optimization uses $k_n=10^3$; continuation uses $k_n$: $100$ (yellow), $200$ (purple), $400$ (green), $800$ (blue), and $10^3$ (red).
}{}}
\label{fg:continuation_plot}
\end{figure}

In this section we have described various ways of handling contacts with Coulomb friction, focusing our discussion on a single contact point.
Now we briefly summarize how these considerations integrate with our differentiable simulation framework.
Recall that for deformable objects the nodes of the FEM mesh directly define the degrees of freedom.
In this case, the contact handling extends naturally from a single point to each mesh node independently.
For rigid bodies, on the other hand, we handle spheres or point sets defining the collision proxy.
The resulting contact forces are then mapped from each contact point to the generalized force vector as described in \S~\ref{sc:details}.

}{}
\rev{
In Fig.~\ref{fg:CylinderDrop}, we compare the different contact models described above through an experiment where a soft cylindrical puck is dropped onto an inclined plane.
We choose a relatively low penalty factor of $k_n = 10^2$ for this example to highlight the differences between soft and hard constraint methods, especially in the static friction phase.
The differences in the normal direction are imperceptible, even with a fairly low penalty factor.
With soft friction constraints the cylinder slides slightly further (Fig.~\ref{fg:CylinderDrop}a) and does not come to a complete stop (Fig.~\ref{fg:CylinderDrop}b) (though the tangential speed is on the order of $1$~mm/s).
Also of note is the fact that the SQP solver converges much slower during the sliding phase (Fig.~\ref{fg:CylinderDrop}c), while our hybrid method performs similar to pure penalty methods in most cases, but delivers equivalent results to the SQP version under static friction.
Conversely, \emph{tanh} penalty forces are highly non-convex around the Coulomb limit, which results in slightly slower convergence during the sticking phase as compared to the linear friction forces.
Overall, the linear penalty approach to friction forces converges faster than \emph{tanh} penalties or hybrid contacts, while the SQP method takes about twice as long in terms of total CPU time; see also Table~\ref{tb:ResultsDeformable}.
}{}


In summary, our validation tests show that penalty-based models of frictional contact approach the ground truth solution defined by complimentary contact constraints and Coulomb's friction law, and they lead to better convergence rates for forward simulation than alternatives based exclusively on hard constraints.
In this context, employing an implicit integration method in conjunction with a line search routine maintains simulation stability at every time step even for very stiff penalties.
Furthermore, our treatment of normal and friction forces makes sensitivity analysis (both the direct and adjoint variation) easy to apply to the simulated motion trajectories.
We note that this would be much more challenging to achieve if we had to take derivatives of the general KKT conditions of the underlying linear complementarity problem.


\rev{
To understand how the different contact models affect the types of inverse problems we aim to solve with our differentiable simulator, we perform another experiment. Here we exhaustively sample the objective function on the task of tossing a ball to a specific target location (see also Fig.~\ref{fg:SphereToPointAndLine}a). In particular, we evaluate the objective function value and its gradients on a regular grid in input parameters $v_x$ and $v_z$ (i.e.~the initial linear velocity in the forward and upward direction, respectively). Note that the ball is initially spinning with a non-zero angular velocity that is kept fixed for all tosses.
Figure~\ref{fg:SphereToPoint-objFcnCombined} illustrates these results.
The two local minima correspond to one-bounce and two-bounce solutions to this control problem.
As can be seen, contacts introduce noise into the objective function (visible as wiggly isolines and somewhat incoherent gradients); this is inevitable as contacts are inherently discontinuous events. Slight changes in the object's initial velocity can lead to a different order in which the nodal degrees of freedom impact the ground. When both the object and the ground are stiff, the noise in the objective function landscape caused by these discretization artefacts can be significant and lead to reduced performance of gradient-based optimization methods. Nevertheless, our experiment shows that the smoothness of the objective function landscape can be effectively controlled through the parameters used by the contact model. This is because smoother contact models enlarge the window of time over which contacts are resolved, and they avoid the use of large impulsive forces. Sensitivities with respect to the exact timing and order of collisions are therefore reduced. This observation, which is supported by the objective function landscapes visualised in Figure~\ref{fg:SphereToPoint-objFcnCombined}, can be exploited to improve convergence rates for the inverse problems that leverage our differentiable simulator.

For the example in Fig.~\ref{fg:continuation_plot}, we evaluate a simple continuation approach. This time around, the task is to throw a geometrically-complex object (a bunny) such that it lands upright in a particular location. When the optimization problem is solved using a stiff contact model, an unfavourable local minimum is quickly reached. In contrast, if the optimization problem starts out with a soft contact model which gets progressively stiffer over time, a much better solution is found.
}{}

\rev{Based on the experimental results described above, we conclude that the linear friction force model offers a favourable trade-off between simplicity, accuracy and practical performance, and as such it is our default choice for the results we present in this paper.}{}

\section{Internal and External Forces in Generalized Coordinates} \label{sc:details}



\rev{
In this section, we describe the models used to generate the forces acting on the multi-body systems simulated within our framework.
We also present basic validation tests of our forward simulation.
}{}

\subsection{Soft bodies}

For deformable elastic objects, we employ a standard Neo-Hookean material model, given by homogeneous Lam\'e parameters $(\mu,\,\lambda)$ and constant mass density $\rho$. As is standard, this material model describes the energy density as a function of the deformation gradient $\mathbf{F}$. Internal shape-restoring forces that arise in response to induced deformations are then computed as the negative gradient of the energy density integrated over each element with respect to the nodal degrees of freedom.

To model the behaviour of real-world objects, the elastic forces described above must be complemented by internal damping forces. 
Most viscosity models, such as the ones described in \cite{Hahn2019}, define the viscous stress (and consequently the damping force) based on the linear strain rate ${({\dot{\mathbf{F}}} + {\dot{\mathbf{F}}^\user1{T}})}$.
One major drawback of these models is that they are not invariant to rotational motion, and therefore damp out the angular velocity of a deformable object during free flight.
\rev{
Brown et al.~\shortcite{Brown2019}, on the other hand, describe a family of rotation invariant viscosity models.
Here, we employ a quadratic model, similar to their power-law damping, defining the viscous stress as a function of the Green strain rate ${\mathbf{D}}$.
}{Instead, we define a viscous stress as a function of the Green strain rate ${\mathbf{D}}$, which is rotation invariant.}
In particular, we define the viscous stress as
\[{{\mbf{\sigma }}_\nu }: = \frac{\nu}{2}  \frac{{d (\operatorname{tr} ({\mathbf{D}}^\user1{T}{\mathbf{D}}))}}{{d \dot{\mathbf{F}}}},\quad {\mathbf{D}}: =   \frac{1}{2}({{\mathbf{F}}^\user1{T}}\dot{\mathbf{F}} + {{\dot{\mathbf{F}}}^\user1{T}}{\mathbf{F}}), \eqn \label{eq:rotInvViscStress} \]
where ${\mathbf{F}}$ is the deformation gradient, $\dot{\mathbf{F}}$ is the velocity gradient, and $\nu$ is the material's viscosity.
We compute this derivative, as well as the corresponding damping matrix entries, using symbolic differentiation.
As this viscosity model is based on a quadratic strain rate, it behaves like a power-law model with flow index $h=2$.

\begin{figure}
\def\svgwidth{\columnwidth}
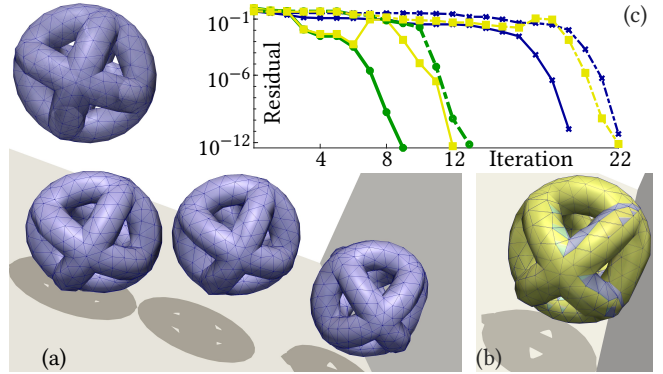
\caption{Dropping an object composed of three tori. Selected frames from the animation (a), and overlay of the wall contact configuration (b) simulated with different contact methods: linear penalty (green), tanh penalty (blue), hybrid (yellow).
Solver convergence (c) for the shown time steps with ground contact (solid), and wall contact (dashed).}
\label{fg:ToriBall-conv}
\end{figure}

Figure~\ref{fg:ToriBall-conv} shows a deformable object composed of three tori in a drop test.
The ground is inclined by $20°$, while the friction coefficient of the ground is $0.4$ and $0.8$ for the wall.
In this example, we use BDF2 time integration.
The rotation invariant viscosity model allows the object to rotate freely in the absence of contacts, but damps out elastic oscillations.
Our three contact methods converge reliably to very low residuals; Fig.~\ref{fg:ToriBall-conv}c shows convergence for two representative time steps during ground and wall contact respectively.
Again, the linear penalty method is fastest, while \emph{tanh} and hybrid contacts are closely matched.
\rev{
These tests further confirm our conclusion that penalty methods are sufficiently accurate when using implicit integration, which allows a high penalty stiffness.
}{}

\subsection{Rigid bodies}

Cartesian-space forces $\mbf{f}$ and torques $\mbf{\tau}$ applied to a rigid body project to generalized coordinates via the standard transformation
\[
{\mathbf{\hat f}} = 
\left( {
\begin{array}{*{20}{c}}
{{{\mathbf{I}}}} & 0\\ 
0 & {\mathbf{J}_{\omega}^\user1{T}}
\end{array}} 
\right)
\left(
{\begin{array}{*{20}{c}}
\mbf{f}\\
{[\hat{\mbf{x}}]\mbf{f} + \mbf{\tau}}
\end{array} 
}\right) \eqn
\label{eq:rbMf},
\]
where the Jacobian $\mathbf{J}_{\omega}$ maps the rate of change of a rigid body’s rotational degrees of freedom to changes in its world-space angular velocity. We use this expression, for example, to apply the contact and friction forces computed in \S~\ref{sc:contact} to rigid bodies that are in contact with the environment. Note that this operation demands the computation of the world coordinates of a contact point, $\mbf{x}(\mbf{q})$, as well as its time derivative $\dot{\mbf{x}}(\mbf{q}) = (\partial\mbf{x}/\partial\mbf{q})  \dot{\mbf{q}}$. 
We parameterize rotations with exponential coordinates $\mbf{\theta}$ and compute derivatives as in \cite{gallego2015compact}.
Force Jacobians, which are needed for both forward simulation and sensitivity analysis, can be easily computed analytically by using the chain rule in conjunction with the derivatives presented in our supplements.

\begin{figure}
\def\svgwidth{\columnwidth}
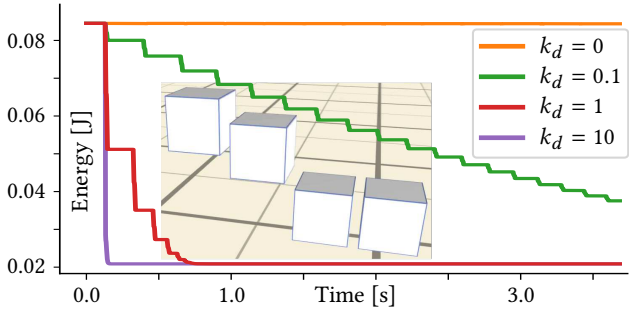
\caption{Total energy over time for bouncing rigid cubes (BDF2) with various ground contact damping coefficients (increasing left to right in the inset image).}
\label{fg:RBcubes-energy}
\end{figure}

One important concept that must still be modelled is the restitution behaviour of rigid body contacts. While post-impact velocities for deformable objects are governed by the material's elastic parameters and internal viscosity, for rigid bodies we must explicitly include a damping force in the normal direction in the event of a contact:
\[
\mathbf{f}_d = -k_d \mathbf N \dot{\mathbf{x}} \quad\text{if } g(\mbf{x}) \leq 0, \quad \mathbf{f}_d = 0 \text{ otherwise}, \eqn \label{eq:RBdamping}
\]
\rev{where $k_d$ is the damping coefficient.
For our implicit soft contacts, this contact damping model replaces the common Moreau impact law used in explicit rigid body engines to model restitution behaviour.
In the absence of external forces, the contact phase for a single one-dimensional point mass $x$ against a wall at $x=0$ can be described as a damped harmonic oscillator
$m\ddot{x} + {k_d}\dot{x} + {k_n}{x} = 0$.
Analysing the exact solution for this oscillator, with initial conditions $x_0=0$ and $\dot{x}_0=-v_\text{in}$, we find the following relation between the damping coefficient and the restitution ratio:
\[\frac{{{v_{{\text{out}}}}}}{{{v_{{\text{in}}}}}} = \exp \left({{ {\frac{{ - \pi {k_d}}}{{\sqrt {4{k_n}m - k_d^2}} }}} }\right), \eqn \label{eq:restitution}\]
where $v_\text{out}$ is the outgoing velocity measured after the first half-period of oscillation.
Restitution occurs only below the critical damping factor, $k_d^2 < 4{k_n}m$.
}{
where $k_d$ is the damping coefficient, and $\mathbf N \dot{\mathbf x}$ is the velocity of the contact point $\mathbf x$ projected onto the normal direction of the signed distance function $g$ describing the contact geometry.}

Figure~\ref{fg:RBcubes-energy} shows a basic test case for our fully implicit rigid body system using BDF2 time integration.
Without additional damping, numerical damping is barely noticeable when time-stepping at $\Delta_t = 1/60$~s. This corresponds to almost perfectly elastic collisions. 
We can effectively control the restitution via our linear contact damping model. 
Note that the symmetry of the contact is maintained over many bounces.
In our video, we also show that rotating the cube slightly to the left quickly breaks this symmetry for comparison.

\subsection{Multi-body systems}

\rev{
As we use implicit integration schemes for time-stepping, we employ (stiff) generalized springs to couple the individual constituents of a multi-body system to each other.
This is a simple, general, and drift-free technique that can be shown to be closely related to Baumgarte-stabilized velocity level constraints for rigid body dynamics~\cite{Catto11}.
In general, for non-dissipative coupling elements, we define a potential energy as a function of points or vectors anchored on the multi-body system.
Taking the gradient of this potential energy with respect to the system's generalized coordinates directly outputs the resulting generalized forces. 
More formally, constraints $\mbf{c}(\mbf{q})$ are enforced through potentials of the form $E(\mbf{q}) = {(k_c/2)}\, \mbf{c}(\mbf{q})^\user1{T}\mbf{c}(\mbf{q})$.

For instance, the constraint
\[{{\mathbf{c}}_s} := \left\| {{\mathbf{x_1}}}(\mbf{q}) - {{\mathbf{x_2}}}(\mbf{q}) \right\| - l_0 = 0 \eqn, \]
asks that a specific distance is maintained between two points on the multi-body system.
Its resulting potential models a stiff linear spring of rest length $l_0$.
We use zero-length springs to formulate ball-and-socket joints.
Furthermore, unilateral springs of (non-zero length), which do not produce a force under compression, model cables and elastic strings (similar to \cite{Bern2019}).

Hinge joints (i.e.~1-DOF revolute joints) connecting two rigid bodies are defined through attachment points ${(\hat{\mathbf{x}}_1,\; \hat{\mathbf{x}}_2)}$ and rotation axes ${(\hat{\mathbf{a}}_1, \; \hat{\mathbf{a}}_2)}$, specified in the local coordinate frame of each rigid body respectively.
We model hinge joints with two constraints: a zero-length spring connecting the attachment points ${(\hat{\mathbf{x}}_1,\; \hat{\mathbf{x}}_2)}$, and another that aligns the rotation axes ${(\hat{\mathbf{a}}_1, \; \hat{\mathbf{a}}_2)}$:
\[{{\mbf{c }}_h} := {{\mathbf{w}}({{{\mathbf{\hat a}}}_2}) - {\mathbf{w}}({{{\mathbf{\hat a}}}_1})} =0, \eqn \]
where $\mbf{w}(\hat{\mbf{a}})$ denotes the mapping from local to world coordinates.

We model active motors by extending hinge joints with a second set of local axes $\hat{\mathbf{b}}_1$ and $\hat{\mathbf{b}}_2$ that are orthogonal to the rotation axes $\hat{\mathbf{a}}_1$ and $\hat{\mathbf{a}}_2$ respectively.
A motor constraint, ${{\mbf{c}}_m}$, enforces a specific relative angle $\alpha$ between $\mbf{w}(\hat{\mathbf{b}}_1)$ and $\mbf{w}(\hat{\mathbf{b}}_2)$:
 \[ {{\mbf{c}}_m}: = \mathbf{w}(\hat{\mathbf{b}}_1) - \mbf{R}(\alpha)\mathbf{w}(\hat{\mathbf{b}}_2). \eqn \]


To model position-controlled motors that are driven by typical Proportional-Derivative controllers, it is also important to add a damping component to the torques these motors generate. To this end we directly define a world-space torque as a function of end-of-time-step angular velocities: \[ 
{{\mbf{\tau }}_{md}}  
:= k_{md} (\mbf{\omega}_1(\mbf{q}, \dot{\mbf{q}}) - \mbf{\omega}_2(\mbf{q}, \dot{\mbf{q}}))
\eqn 
\]
which we then project into generalized coordinates using Eq.~\ref{eq:rbMf}.
}{}







\section{Results}

%
%

We now turn our attention to \emph{optimization problems} based on our dynamics system.
In this section, \rev{we show the effectiveness of our differentiable simulation}{before demonstrating its usefulness} for gradient-based optimization \rev{compared to gradient-free alternatives}{} on various examples\rev{.}{, and comparing the results to real-world experiments.}
\rev{We present results on various applications: material parameter estimation including contacts, optimizing initial conditions, machine learning with our differentiable simulator directly built into the loss function, and trajectory optimization for robotics.}{}
Unless stated otherwise we use a penalty factor of $k_n = 10^3$ and convergence tolerance $\varepsilon = 10^{-10}$.

\subsection{Material parameter estimation} \label{sc:paramEst} 

\rev{}{ 
\begin{figure*}
\def\svgwidth{\textwidth}
\input{img/noiseResults.pdf_tex}
\caption{Errors obtained when optimizing for ground truth initial conditions and material parameters on synthetic data for linear penalty contacts (a), tanh penalties (b), and hybrid contacts (c).
We show relative errors for Young's modulus $E$ and the coefficient of friction $c_f$, and absolute errors for initial conditions (position and velocity).}
\label{fg:SphereMarkersGroundTruth}
\end{figure*}
}
\rev{
Our system allows us to estimate material parameters such as stiffness and damping of deformable objects.
We capture the real-world behaviour of our specimens using either an optical motion capture system, or a \emph{Kinect v2} depth camera.

In the former case, we track up to six labelled optical markers on the specimen at a frame rate of $120$~Hz using an array of $10$ \emph{OptiTrack Prime 13} cameras.
The system calibration ensures that the world-space coordinates align with the ground and wall planes in order to include these rigid obstacles in the simulation.
For motion capture data, the objective function measures the sum of squared distances between the tracked marker position and the corresponding location on the simulated mesh for all time steps.
In each time step, we only consider markers that are currently visible to the tracking system.

In the latter case, we read 3D point clouds from the \emph{Kinect} at $30$~Hz and then identify the ground and wall planes in a manual post-processing step.
We apply box filters in both world and colour space to identify which points correspond to the surface of the specimen.
As we do not have a direct correspondence between tracked points and mesh locations in this case, the objective function instead measures the sum of all filtered points to their respective closest point on the surface of the simulated mesh.
}{
We present an application of our system to material parameter estimation based on motion capture data, tracking optical markers on the real-world specimen at a frame rate of $120$~Hz.
Our motion capture system consists of $10$ \emph{OptiTrack Prime 13} cameras.
The recorded data contains a motion of the specimen being thrown onto a table.
}

\rev{For parameter estimation, we employ a continuation strategy in time, first}{We start by} optimizing the initial conditions of the simulation (in terms of position, orientation, velocity, and spin) to match the recorded motion during the ballistic phase (before the first contact).
In the second phase, we keep the initial conditions fixed, and optimize the material parameters such as to best approximate the first bounce of the recorded motion.
Finally, we add a third phase where all parameters (material and initial conditions) are optimized simultaneously for the entire motion.
\rev{
In our experiments, we find that L-BFGS is well suited for these optimization tasks.
}{The first and third phase use L-BFGS only, whereas the middle phase (where we optimize for material parameters) starts with a relatively short ADAM run of at most $75$ iterations, and then uses L-BFGS starting from the best previously found parameters.

We evaluate the robustness of this approach on synthetic ground truth data containing only a single bounce in Fig.~\ref{fg:SphereMarkersGroundTruth}.
Starting from a different initial guess, about $10$ times softer than the ground truth material, the optimization should recover both the initial conditions and the material parameters while we add uniform random noise to the ground truth data.
Both penalty based approaches find very accurate solutions in the absence of noise, and provide good approximations even at higher noise levels.
In contrast, the hybrid method, where static friction is enforced by hard constraints, causes the optimization to find unfavorable local minima even in the absence of noise.
In particular, the coefficient of friction is not recovered correctly at increased noise levels.
Consequently, we prefer penalty based methods for optimization tasks, but we still show examples where the hybrid method can be utilized for optimization.
}

\begin{figure}
\def\svgwidth{\columnwidth}
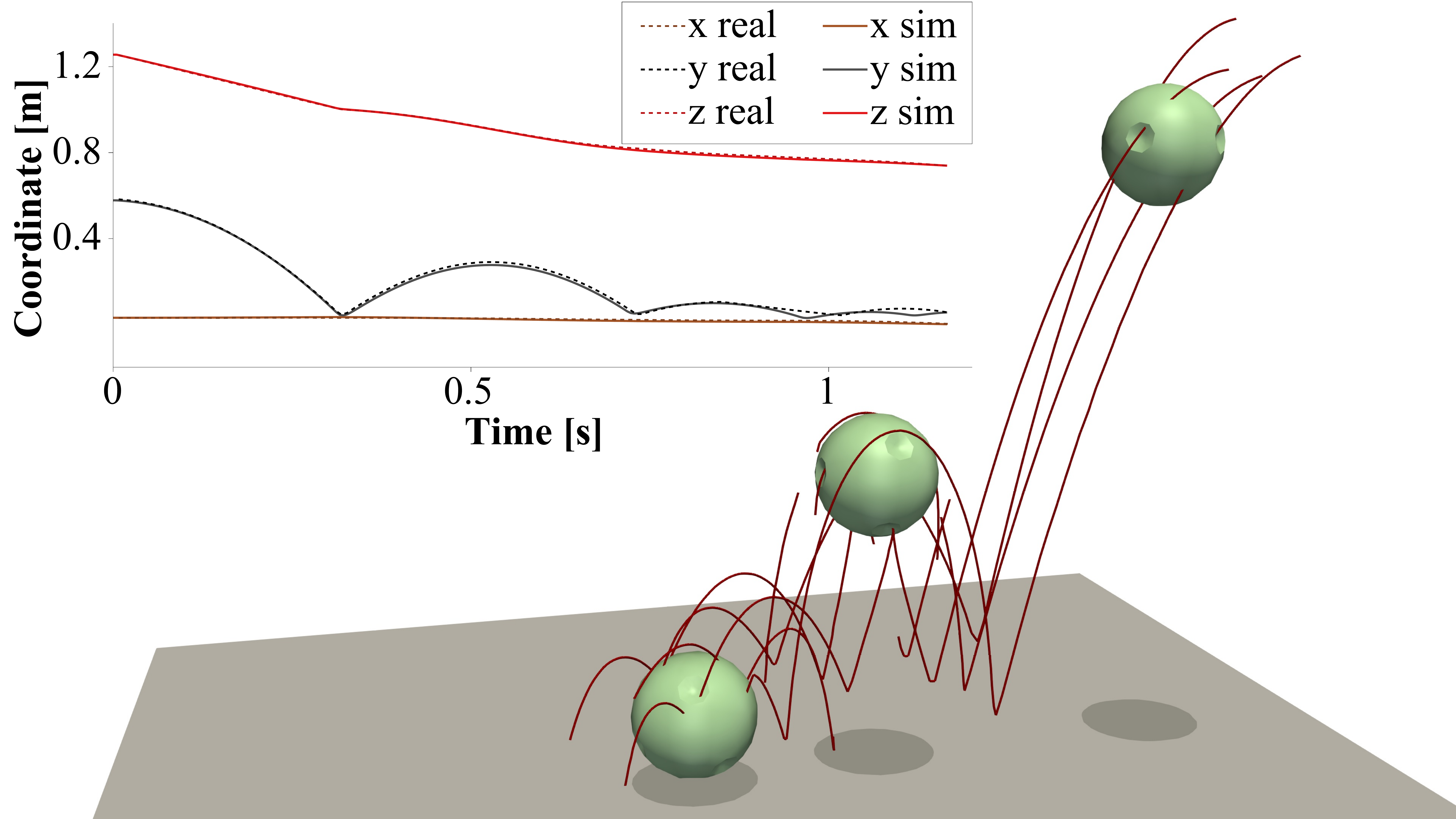
\caption{Parameter estimation for throwing a sphere. The 3D image shows motion capture trajectories for all six markers (two along each coordinate axis) and snapshots of the best fitting simulation. The inset graph shows captured and simulated trajectories for the front facing marker.}
\label{fg:SphereMarkers}
\end{figure}

\begin{figure}
\def\svgwidth{\columnwidth}
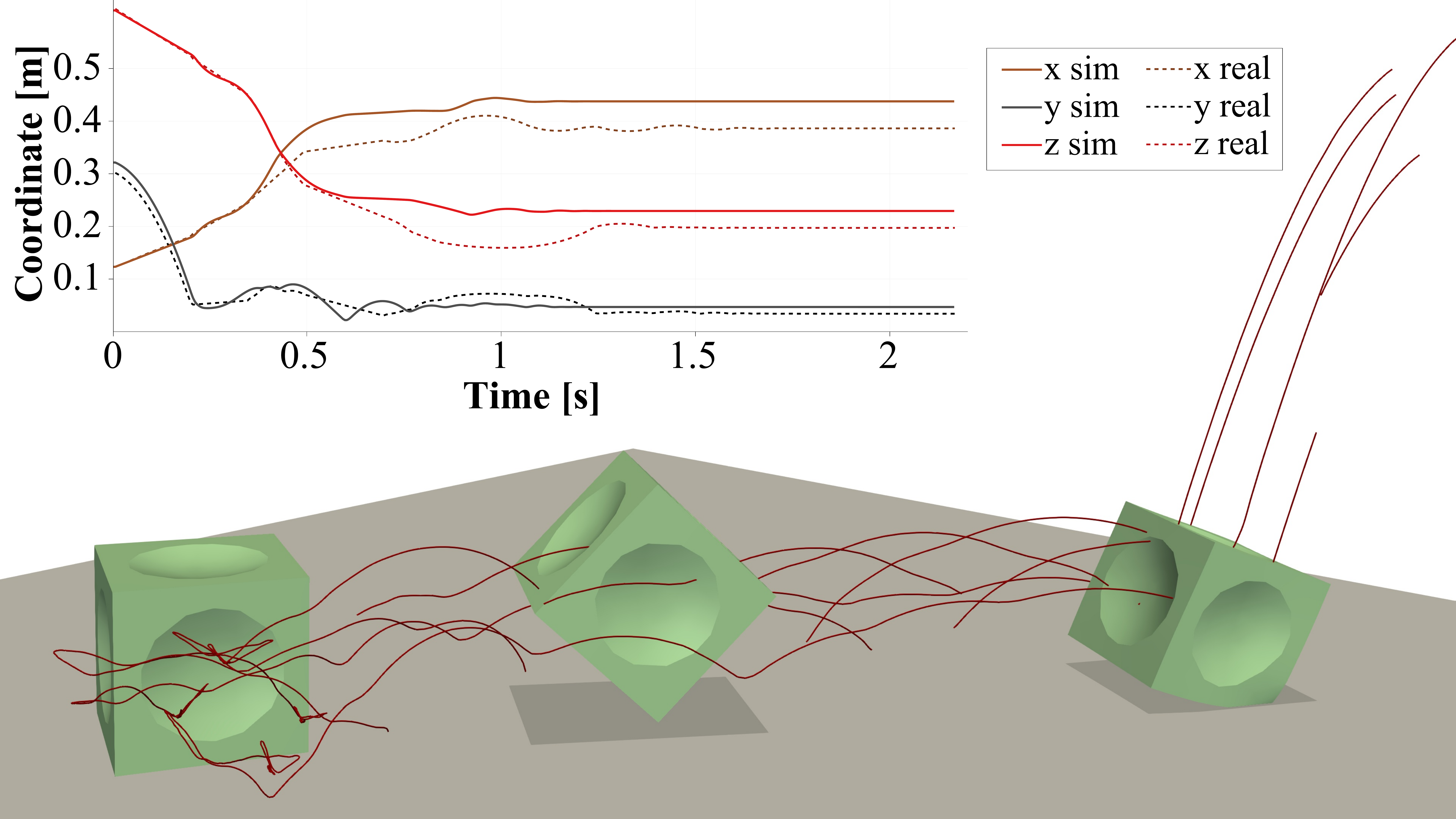
\caption{Parameter estimation for throwing a cube. The 3D image shows motion capture trajectories for all six markers (one in the centre of each face) and snapshots of the best fitting simulation. The inset graph shows captured and simulated trajectories for the front facing marker.}
\label{fg:CubeMarkers}
\end{figure}

\begin{figure}
\def\svgwidth{\columnwidth}
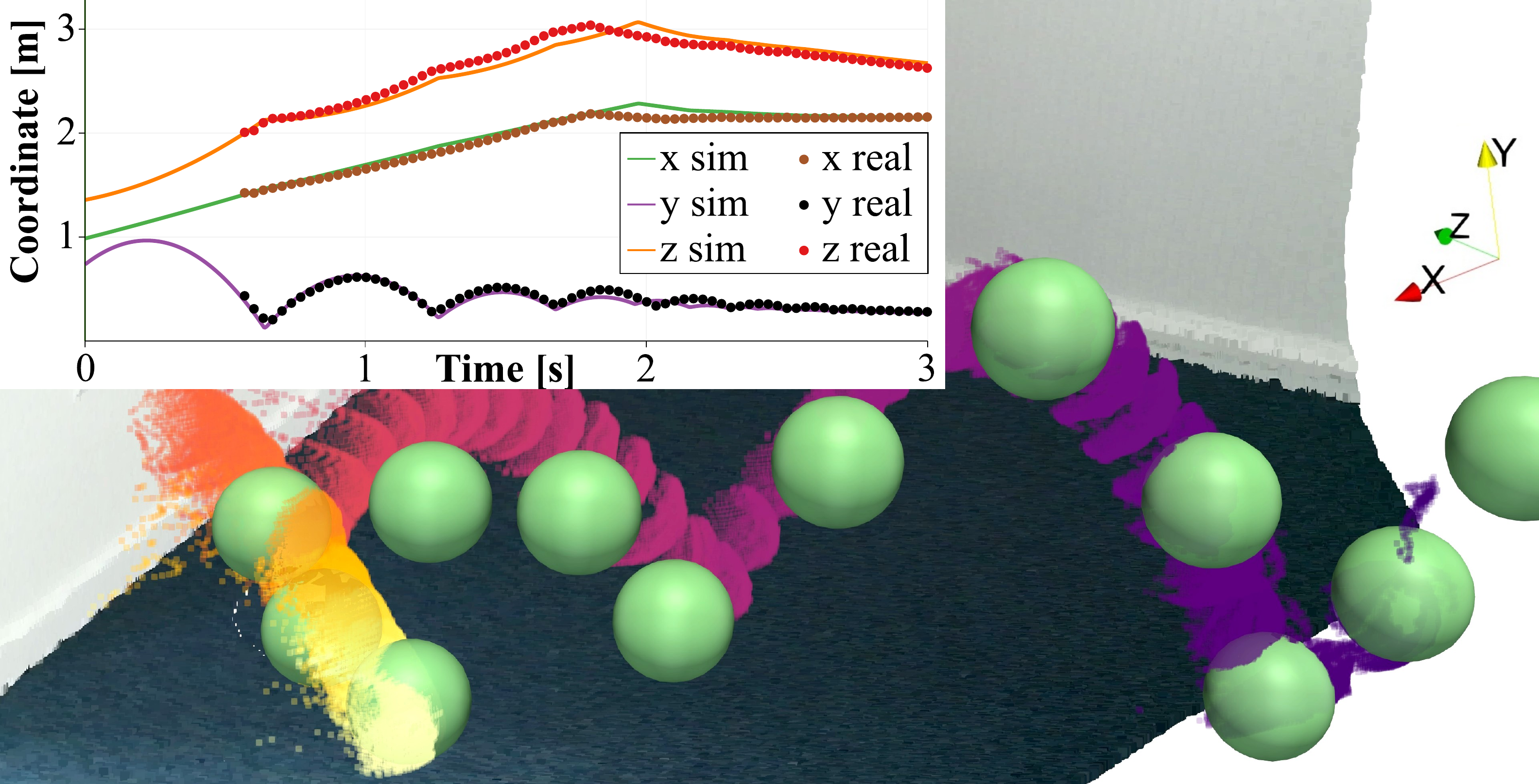
\caption{\rev{Parameter estimation with Kinect data. Image shows input point cloud (time colour-coded purple to yellow) and representative time steps of the simulation result (green). Graph shows average coordinates; axes represent the Kinect camera orientation (z forward, y up).}{}}
\label{fg:KinectBall}
\end{figure}

\rev{Here, we show three}{We show} results for material parameter estimation \rev{of}{for two} real-world specimens\rev{.}{,}
\rev{We prepare two custom-made elastic foam specimens,}{} a sphere and a cube\rev{, for motion capture}{} with six slightly inset motion capture markers each\rev{, see Fig.~\ref{fg:SphereMarkers} and \ref{fg:CubeMarkers}.}{,}
\rev{The motion capture data provides direct correspondences between the tracked markers and their simulated counterparts, allowing us to find the initial orientation and angular velocity at this early stage.}{}
\rev{In the first example, Fig.~\ref{fg:SphereMarkers}, we then optimize material parameters for the duration of the first bounce, and finally all parameters over the entire recorded trajectory.
The second example, Fig.~\ref{fg:CubeMarkers}, uses a more automated approach, optimizing all parameters over increasing time horizons.
Apart from the material parameter optimization in the first example, which includes a short ADAM phase, we use L-BFGS for all these optimizations.}{}
In our accompanying video we also show verification tests for both of these results, where we use the material parameters obtained via these optimizations, and then only fit the initial conditions to the ballistic  phase of a different recorded motion.
Please also refer to Table~\ref{tb:ResultsDeformable} for details on material parameters and runtime.

\rev{Finally, we show an example for a foam ball without additional markers, where we record the real-world motion with a \emph{Kinect} depth camera, Fig.~\ref{fg:KinectBall}.
In this case, we minimize the distance from the simulated surface to the recorded point cloud data, which means that we do not have any rotational information about the real-world specimen.
Nevertheless, by allowing the optimization to change initial conditions during later stages where contacts are taken into account, we find a good match between real and simulated motion.
We employ the same approach to estimate parameters of a common tennis ball, which we subsequently throw with a robot as discussed in the next section.
}{}

\subsection{Throwing}

\begin{figure}
\def\svgwidth{\columnwidth}
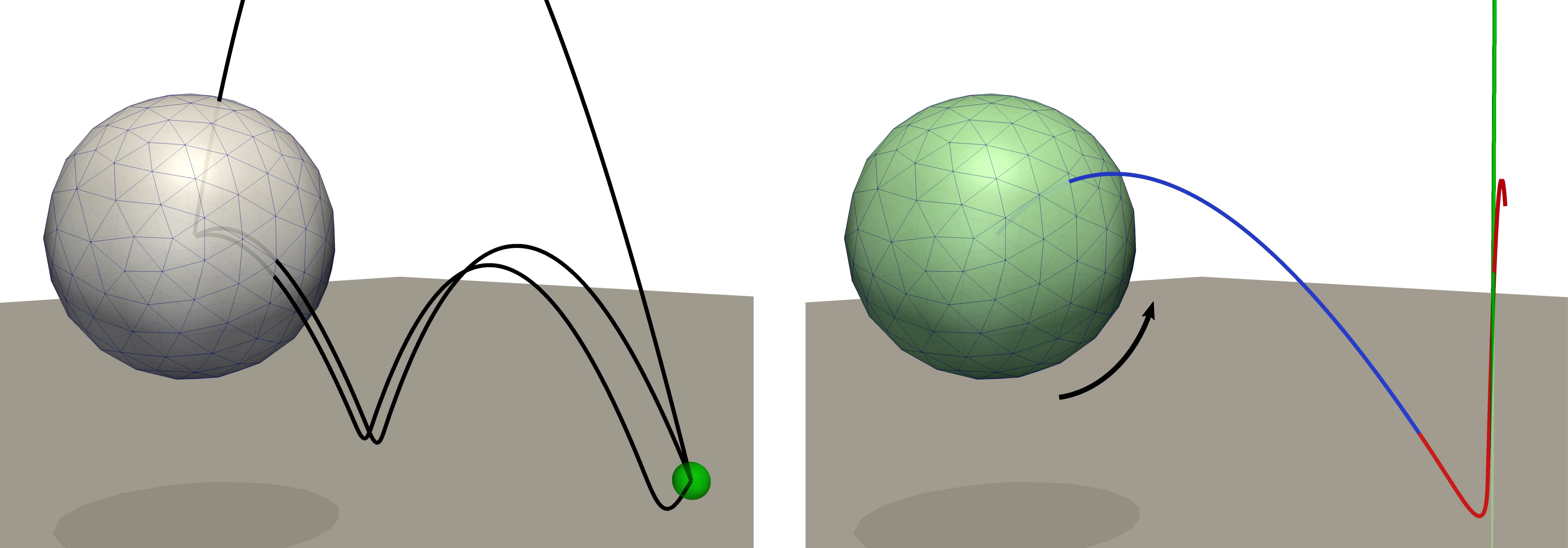
\caption{Throwing a deformable ball: a point target for the ball's centre of mass (a) admits multiple exact solutions with either zero, one, or two bounces off the floor.
Asking the second half of the c.o.m. trajectory (red) to be as close to a vertical line as possible (b) requires a trade-off between forward motion and back-spin (black arrow) such that friction slows the ball down when it bounces off the ground.}
\label{fg:SphereToPointAndLine}
\end{figure}

We can parametrize, and optimize for, the initial conditions of our simulation, such as in the examples shown in Fig.~\ref{fg:SphereToPointAndLine}.
In these cases, we must account for the contribution of the initial conditions to the objective function gradient in Eq.~\eqref{eq:adjGrad}.
\rev{
While previous work provides an adjoint formulation for general, implicitly defined, initial conditions \cite{Bradley2013}, when directly parametrizing initial conditions we find it more convenient to calculate the corresponding derivatives explicitly.

Parameters that define initial conditions, ${{\mathbf{p}}_0}$, are parameter variables that affect only the initialization of the time integration scheme, but do \emph{not} directly affect any of the \emph{unknown} states ${\mathbf{\tilde q}}$.
We can therefore compute the derivatives of the residuals w.r.t.~these parameters analytically: $\partial {{\mathbf{r}}^i}/\partial {{\mathbf{p}}_0} = (\partial {{\mathbf{r}}^i}/\partial {\mbf{\chi }})(\partial {\mbf{\chi }}/\partial {{\mathbf{p}}_0})$, where ${\mbf{\chi }}$ refers to the initial state of the time integrator.
The first term follows directly from the choice of time integration method, while the second term follows from the parameterization of initial conditions.
Finally, these derivatives are added to the matrix ${\mathbf{\tilde A}}$ of the sensitivity system, where each block now becomes ${{\mathbf{A}}^i}: = \partial {{\mathbf{r}}^i}/\partial {\mathbf{p}} + \partial{{\mathbf{r}}^i}/\partial{{\mathbf{p}}_0}$.
Note that only the first few time steps receive a non-zero update, depending on the chosen time-discretization scheme.
}{
While previous work provides an adjoint formulation for general, implicitly defined, initial conditions \cite{Bradley2013}, when directly parametrizing initial conditions it is usually more convenient to calculate the derivatives of the initial state wrt.~these parameters directly.
In particular, we compute ${\mathbf{s}}_q^0 = d{\mathbf{q}^0}/d{{\mathbf{p}}_0}$, where ${{\mathbf{p}}_0}$ refers to those parameters that directly define initial conditions (such as position, orientation, velocity, or spin at $t=0$).
Adapting the formulation of \cite{Hahn2019} to this case, we add the terms corresponding to the initial conditions to Eq.~\eqref{eq:adjGrad}, which yields
\[\frac{{d\Phi }}{{d{\mathbf{p}}}} = \frac{{\partial \Phi }}{{\partial {\mathbf{p}}}} - {{\mbf{\tilde \lambda }}^\user1{T}}{\mathbf{\tilde A}} + {\mbf{\lambda }}_0^\user1{T}{\mathbf{s}}_q^0 + \dot{\mbf{\lambda }}_0^\user1{T}{{\mathbf{M}}^\user1{T}}\dot{\mathbf{s}}_q^0,\]
where ${\mbf{\lambda }}_0$ refers to the adjoint state corresponding to the start of the simulation, and time derivatives are again approximated according to the time discretization scheme of our choice.
}

In our first tests, we optimize initial linear and angular velocities for throwing a deformable ball, Fig.~\ref{fg:SphereToPointAndLine}.
The objective function measures the distance from the ball's centre of mass to a specific target point at the end of the simulation (Fig.~\ref{fg:SphereToPointAndLine}a), or to a target line over a specified time range (Fig.~\ref{fg:SphereToPointAndLine}b), respectively.

\begin{figure}
\def\svgwidth{\columnwidth}
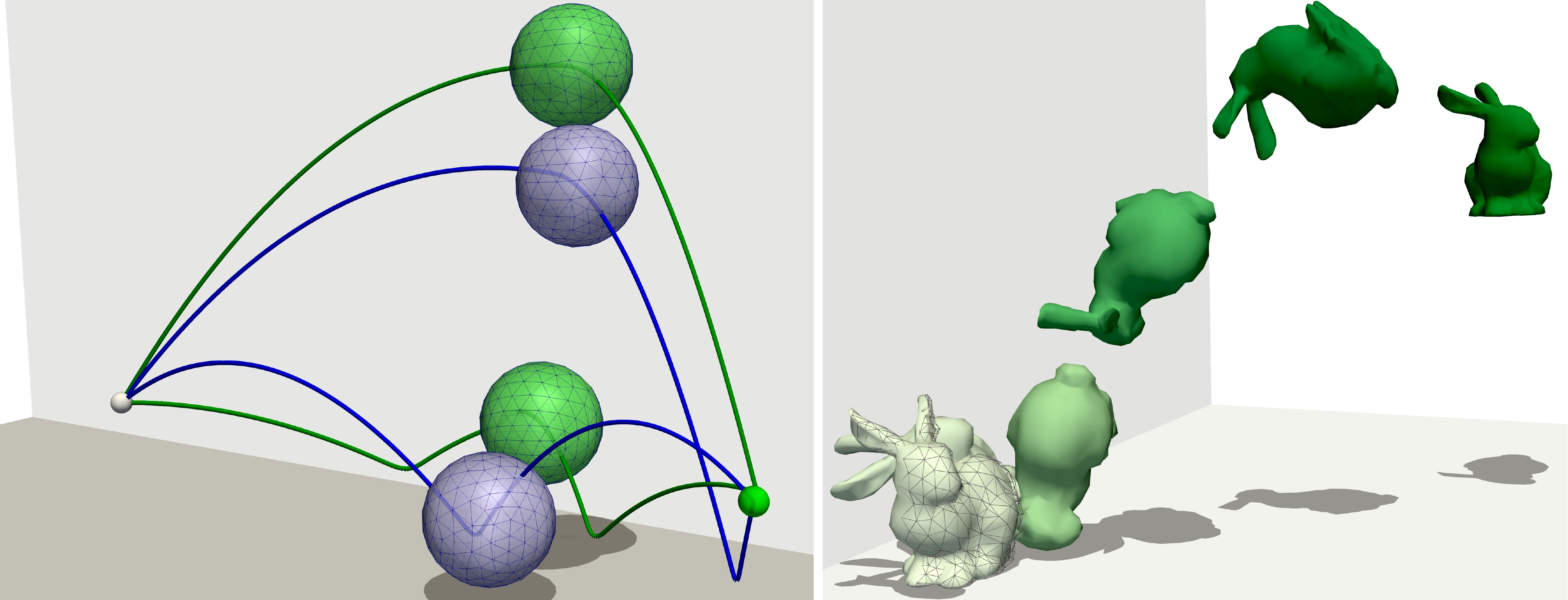
\caption{\rev{
Throwing with multiple contacts: (a) multiple paths for the ball's c.~o.~m. to reach the target point, labelled by contact sequence (`w'~wall, `f'~floor);
(b) throwing the bunny to a specific target pose (wireframe) after bouncing off the floor and the wall (time colour-coded from dark to bright).
}{Optimizing initial linear and angular velocity for throwing a bunny (a) such that it lands on its feet after bouncing off the floor and the wall. Transparent shape shows the initial guess, which falls over (solid mesh shows the final result), and (b) throwing a ball into a cup via multiple bounces.}}
\label{fg:SphereAndBunnyThrow}
\end{figure}

\rev{
In order to compare our results to a gradient-free sampling method, we run CMA-ES on the optimization problem in Fig.~\ref{fg:SphereToPointAndLine}a.
Qualitatively, gradient-based approaches are less likely to traverse a saddle point, whereas sampling methods explore the parameter space more randomly in the early stages before settling into a local minimum.
In this particular case, using L-BFGS and linear penalty contacts requires $28$ simulation runs to find an exact solution for the two-bounce motion (relative objective function value ${\Phi/\Phi_0<10^{-25}}$) in a matter of minutes (Table~\ref{tb:ResultsDeformable}), while CMA-ES requires $1042$ simulations and $2$h~$15$m to find an approximate solution with ${\Phi/\Phi_0 \approx 10^{-6}}$.

On a more complex example (Fig.~\ref{fg:SphereAndBunnyThrow}b), L-BFGS finds a better solution (${\Phi/\Phi_0 \approx 4\cdot10^{-4}}$) after $319$ simulations, whereas CMA-ES returns a noticeably worse result (${\Phi/\Phi_0 \approx 10^{-2}}$) even after running over $8000$ simulations. 
We observe the same behaviour for a trajectory optimization test, similar to Fig.~\ref{fg:UR5-tests}, where CMA-ES fails to produce an acceptable solution after multiple hours, whereas our system yields a good result in a few minutes using direct sensitivity analysis and Gauss-Newton optimization; see our supplements for details.
}{}

\rev{We then show results for artistic control of animations.
In particular, we find optimal throwing velocities such that an elastic object hits a specified target after multiple bounces.
We first extend the example of Fig.~\ref{fg:SphereToPointAndLine}a by including a wall and increasing the distance to the target, Fig.~\ref{fg:SphereAndBunnyThrow}a.
Depending on the initial conditions, we can now find multiple paths to the target with various bounce patterns, as labelled in the image.
In these examples, using smoother \emph{tanh} friction forces (blue) yields slightly better results than linear ones (green).

We can also throw the Stanford bunny (again including a contact with a wall) such that it lands at a specific target location (Fig.~\ref{fg:SphereAndBunnyThrow}b).
In this example, the objective function measures the squared distance to the target pose for each mesh node, and also includes a regularizer that additionally penalizes solutions where the bunny falls over.}{}

\rev{
After performing parameter estimation for a tennis ball, as described in the previous section, we also optimize initial conditions for a new throw such that the tennis ball hits a specific location on the wall after bouncing off a table once.
For the resulting initial position and velocity, we then generate a throwing motion using a standard inverse kinematics model for a UR5 robot, as shown in our video.
}{}

\subsection{Self-supervised learning of control policies}

\begin{figure}
\def\svgwidth{0.9 \columnwidth}
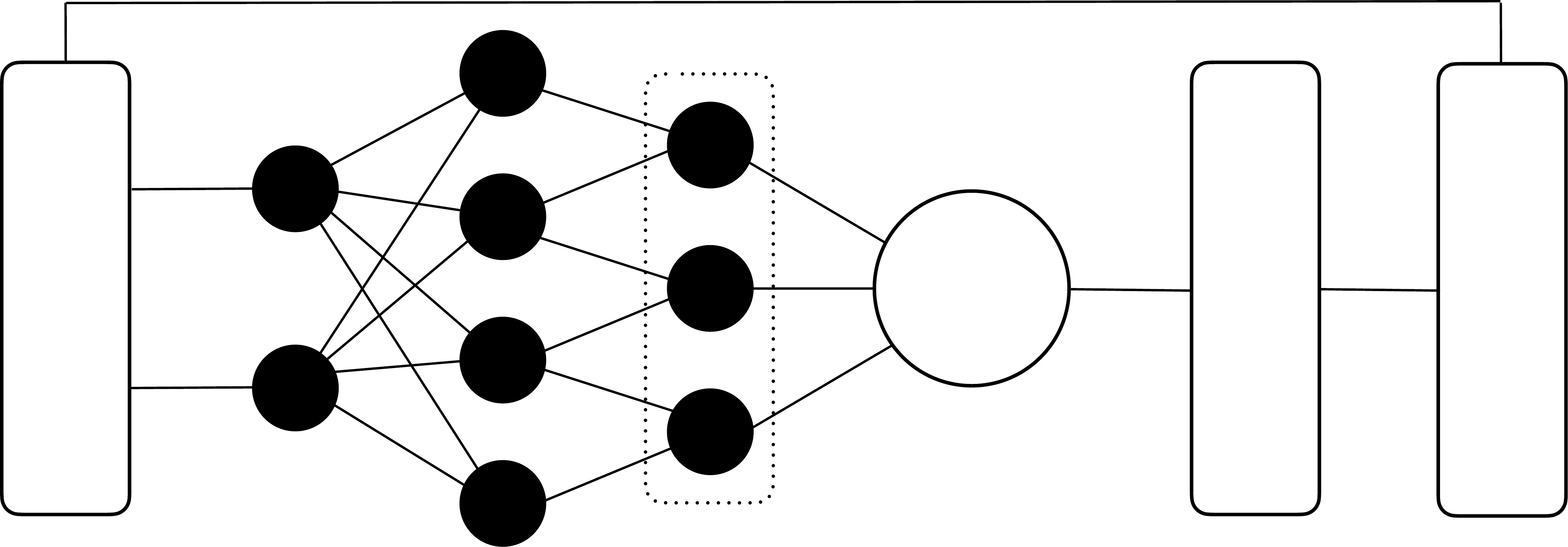
\caption{Integrating differentiable simulation with a neural network.} \vspace{-8pt}
\label{fg:MLschematics}
\end{figure}

Learning-based methods that leverage neural networks and simulation data to train control policies have achieved impressive results for various control applications. Forward simulation is commonly used as an infinite data source that is sampled over initial conditions and control parameters to generate training data. While this data-driven approach effectively decouples learning from simulation and thus simplifies implementation, it critically relies on the parameter space sampling reflected in the training data to yield an appropriate coverage in performance space. 
\rev{We}{In the following example, we} pursue an alternative strategy that, rather than using a fixed training set, integrates simulation directly into the loss function,\rev{ as illustrated in Fig.~\ref{fg:MLschematics},}{} thus enabling the learning algorithm to exploit the map between parameter and performance space provided by our differentiable simulator.

\rev{
As in the other applications, the objective (or \emph{loss}) function measures the simulation result $\mbf{\tilde q}$ against a desired target behaviour $\mbf{q}^*$, i.e.~$\Phi( \mbf{q}^*, \mbf{\tilde q})$.
We then train the neural network ${\phi  ({\mathbf{q}}^*,{\mathbf{w}})}$ to return simulation parameters $\mbf{p}$ that achieve the given target.
The result is a weight vector $\mbf{w}$ for the network that minimizes the training loss:}{}
\begin{equation}
    \min_{\mbf{w}} \frac{1}{n} \sum\nolimits_{l=1}^n \Phi(\mbf{q}_l^*, {\mathbf{\tilde q}}(\phi ({\mathbf{q}}_l^*,{\mathbf{w}}))).
\end{equation}
\rev{Note that the simulated trajectory is a function of the parameters returned by the neural network, i.e.~${\mathbf{\tilde q}}(\phi ({\mathbf{q}}^*,{\mathbf{w}}))$.
Consequently, a \emph{differentiable} simulation is key to computing gradients during training while avoiding costly finite differencing.}{}
We demonstrate this approach on a simple game where the task is to find the throwing velocity for a rigid ball such that, after a single impact with the ground, it hits a given target position.
To this end, we train a neural network to return the throw velocity
that, when used for simulation, yields a trajectory \rev{approximately terminating at}{that closely approximates} the target position.
\rev{}{The simulation returns all positions $\{\mathbf{q}_i\}$ and velocities $\{\dot{\mathbf{q}}\}$, which we use to define the sample loss as
\begin{equation}
    \Phi( \mbf{x}^*, \mbf{\tilde x}) = S(\mbf{x}^*, \mbf{\tilde x}) + k_p \, P(\mbf{\tilde x}) \ ,
\end{equation}
where $P$ is a penalty term that discourages solutions without contact and $S$ is an objective term measuring the distance to the target. More specifically, we} 
\rev{The objective (or loss) function}{} measures proximity \rev{to the target location}{} using a soft minimum over the descending part of the trajectory\rev{, and includes a penalty term that discourages solutions without contact.}{ $\mathbf{q}^{c_j}$.}
Rather than measuring distance at a specific time, this approach provides more flexibility in terms of timing, allowing the learning algorithm to find better solutions. 

\begin{figure}
\def\svgwidth{\columnwidth}
\input{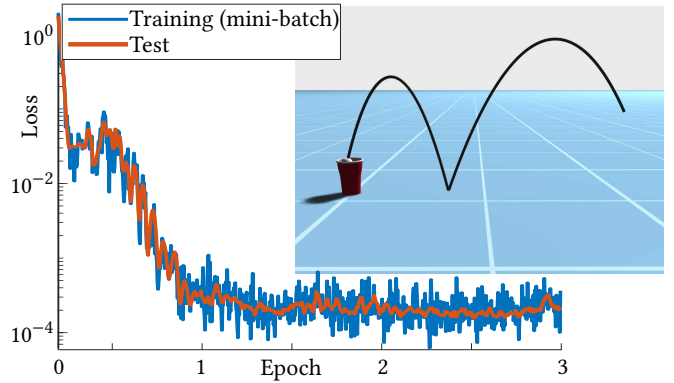}
\caption{\rev{
Convergence of the learning process for the throw controller (left) and the corresponding network architecture (right). The neural network outputs the initial velocity of the simulation such as to hit the given target after one bounce.
}{}}
\label{fg:beerPong}
\end{figure}

We select $n=1000$ target positions for training and $100$ for testing, uniformly sampled in a rectangular region. We train using ADAM \cite{Kingma2014} with $\beta_1=0.95$, $\beta_2=0.999$, $\varepsilon=10^{-8}$, and a mini-batch size of 5. We start with a learning rate of $10^{-2}$, which is reduced by a factor of $0.5$ after each epoch\rev{. On average, each epoch takes about $450$~s of CPU time to compute.}{, which took $447s$ to simulation on average.}
The architecture of the network is shown in Fig. \ref{fg:beerPong}. 
It is worth noting that, even in this comparatively simple example, accounting for friction in the simulation is crucial for accurate control\rev{;}{: as illustrated in Fig.~\ref{fg:beerPong},} a controller trained in a friction-less environment will systematically fail to hit the target.

\rev{}{
\begin{figure}
    \centering
    \includegraphics{img/beer_pong.png}
    \caption{Impact of friction for learning-based control. Using the output from a controller trained in a friction-less environment, the ball will miss the target after frictional contact with the ground (red). Accounting for friction during training (black) leads to throw velocities that reliably hit the target.}
    \label{fig:learningFrictionImpact}
\end{figure}

\begin{figure}
    \centering
	\includegraphics[width=\columnwidth]{img/beer_pong_combined.pdf}
    \caption{Convergence of the learning process for the throw controller (left) and the corresponding network architecture (right).}
    \label{fig:beer_pong_architecture}
\end{figure}
}

\subsection{Trajectory optimization}
 
\rev{We present various applications to robot control using trajectory optimization.
}{Trajectory optimization is a widely adopted tool in control, and our}
\rev{Our }{}contact-aware differentiable simulation is well-suited for these applications.
\rev{We optimize for }{We assume }per-time-step control parameters 
$\mathbf{p}^i$ 
\rev{representing either the position and orientation of a robotic end-effector, or target motor angles.
Note that each subset of parameters affects only one time step, while the entire parameter vector $\mbf{p}$ remains formally time independent.}{}

\rev{We manually define target trajectories for specific feature points.
The objective function $\Phi$ again measures squared distances between the simulated and target trajectories of these features.
}{The objective $\Phi$ is set to the sum of squared differences between parameterized locations on our multibody assemblies, and corresponding targets.}
To ensure temporal \rev{smoothness}{coherence} of parameters, we add the \rev{regularization}{smoothing} term $\beta \sum_i ||\mathbf{p}^{i+1} - \mathbf{p}^i||^2$ to $\Phi$.

\begin{figure}
\def\svgwidth{\columnwidth}
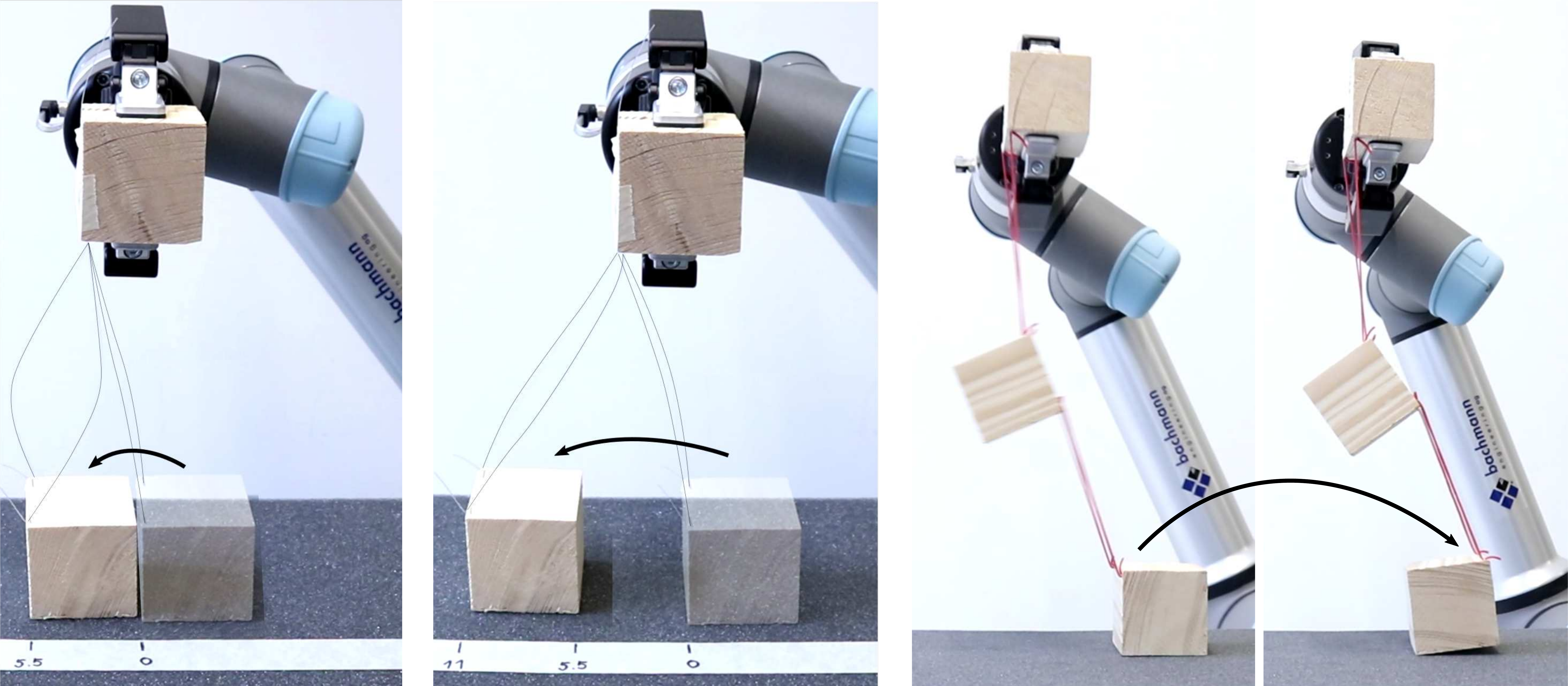
\caption{\rev{
Optimizing end-effector trajectories for robotic control of coupled dynamic systems.
Our robot drags a rigid cube attached with nylon strings over a distance of $5$~cm (a) or $11$~cm (b) respectively,
and actuates a coupled system composed of two rigid cubes and four elastic rubber bands (c) such that the lower cube is tipped over.
}{Our differentiable simulator can be used for a variety of control applications in robotics, as shown with these two experiments. a) An optimal motion trajectory is generated for a physical robot, allowing it to precisely move a wooden block to a prescribed location; this wooden block is connected to the robot's end effector via strings. b) A legged robot equipped with compliant actuators is able to learn how to walk effectively in a simulation environment.}}
\label{fg:UR5-tests}
\end{figure}

We first optimize the trajectory for a 6-DOF robotic end-effector over a time interval of $1$~s, and test our results on a real-world \emph{Univeral Robotics UR5} robot running at a controller frequency of $125$~Hz, 
Fig.~\ref{fg:UR5-tests}.
This test shows that we can effectively optimize for a large number of parameters, as each time step ($\Delta_t = 1/60$~s) has its own set of end-effector coordinates.
We also demonstrate that our simulated results carry over to the real world by having the robot perform the resulting motion repeatedly for comparison (see our video).

\rev{}{
\begin{figure}
\def\svgwidth{\columnwidth}
\input{img/Robot-drag-and-Quadruped.pdf_tex}
\caption{Our differentiable simulator can be used for a variety of control applications in robotics, as shown with these two experiments. a) An optimal motion trajectory is generated for a physical robot, allowing it to precisely move a wooden block to a prescribed location; this wooden block is connected to the robot's end-effector via strings. b) A legged robot equipped with compliant actuators is able to learn how to walk effectively in a simulation environment.}
\label{fg:Robot-drag-and-Quadruped}
\end{figure}
}

\rev{
Another example using our differentiable simulator in a trajectory optimization setting is the manipulation of a sheet modelled as a mass-spring system, Fig.~\ref{fg:teaser}. The control parameters are the positions of two handles each attached at a corner of the sheet. Initially, the sheet lies on the floor facing upward. We then implement the following objectives as target states on the point masses: 
At time $t=1$~s, we ask the sheet to be flipped facing downwards, whereas at time $t=2$~s we ask the sheet to be flipped back to face upwards and moved to the right. Adding a smoothness regularizer for the control parameters, the trajectory optimization finds a solution after a few Gauss-Newton iterations, as can be seen in the accompanying video. 
}{}

In our most complex example, we optimize for control inputs of a legged robot actuated by compliant motors.
For this experiment, we model the robot's actuators as PD controllers with relatively low gains, a reasonable model of position-controlled motors, implemented as soft, damped angular constraints between the coordinate frames of adjacent rigid links.
We use the motion synthesis tool of~\cite{Geilinger2018} to create a nominal motion trajectory for this robot.
However, their trajectory optimization model is based on a relatively coarse approximation of the robot's dynamics and assumes precise and strong actuators.
Unsurprisingly, due to these modelling simplifications, when our compliant robot is attempts to track the planned nominal trajectory, it fails to locomote effectively. 

We note that this use-case is motivated by real-world challenges. Compliance, whether parasitic (e.g. motors that are not infinitely strong) or purposefully built in (e.g. rubber feet designed to soften impacts), is a defining characteristic of physical robots. Most existing motion planning and trajectory optimization algorithms, however, are not able to account for it. Our differentiable simulator on the other hand allows us to optimize the robot's motion by directly considering its full body dynamics, including the compliant nature of its actuators and feet.
The result, as shown in the accompanying video, is a successful locomotion gait for this robotic creature with compliant actuators and soft feet.

\begin{figure}
\def\svgwidth{\columnwidth}
\begingroup%
  \makeatletter%
  \providecommand\color[2][]{%
    \errmessage{(Inkscape) Color is used for the text in Inkscape, but the package 'color.sty' is not loaded}%
    \renewcommand\color[2][]{}%
  }%
  \providecommand\transparent[1]{%
    \errmessage{(Inkscape) Transparency is used (non-zero) for the text in Inkscape, but the package 'transparent.sty' is not loaded}%
    \renewcommand\transparent[1]{}%
  }%
  \providecommand\rotatebox[2]{#2}%
  \newcommand*\fsize{\dimexpr\f@size pt\relax}%
  \newcommand*\lineheight[1]{\fontsize{\fsize}{#1\fsize}\selectfont}%
  \ifx\svgwidth\undefined%
    \setlength{\unitlength}{1670.7881513bp}%
    \ifx\svgscale\undefined%
      \relax%
    \else%
      \setlength{\unitlength}{\unitlength * \real{\svgscale}}%
    \fi%
  \else%
    \setlength{\unitlength}{\svgwidth}%
  \fi%
  \global\let\svgwidth\undefined%
  \global\let\svgscale\undefined%
  \makeatother%
  \begin{picture}(1,0.37325022)%
    \lineheight{1}%
    \setlength\tabcolsep{0pt}%
    \put(0,0){\includegraphics[width=\unitlength,page=1]{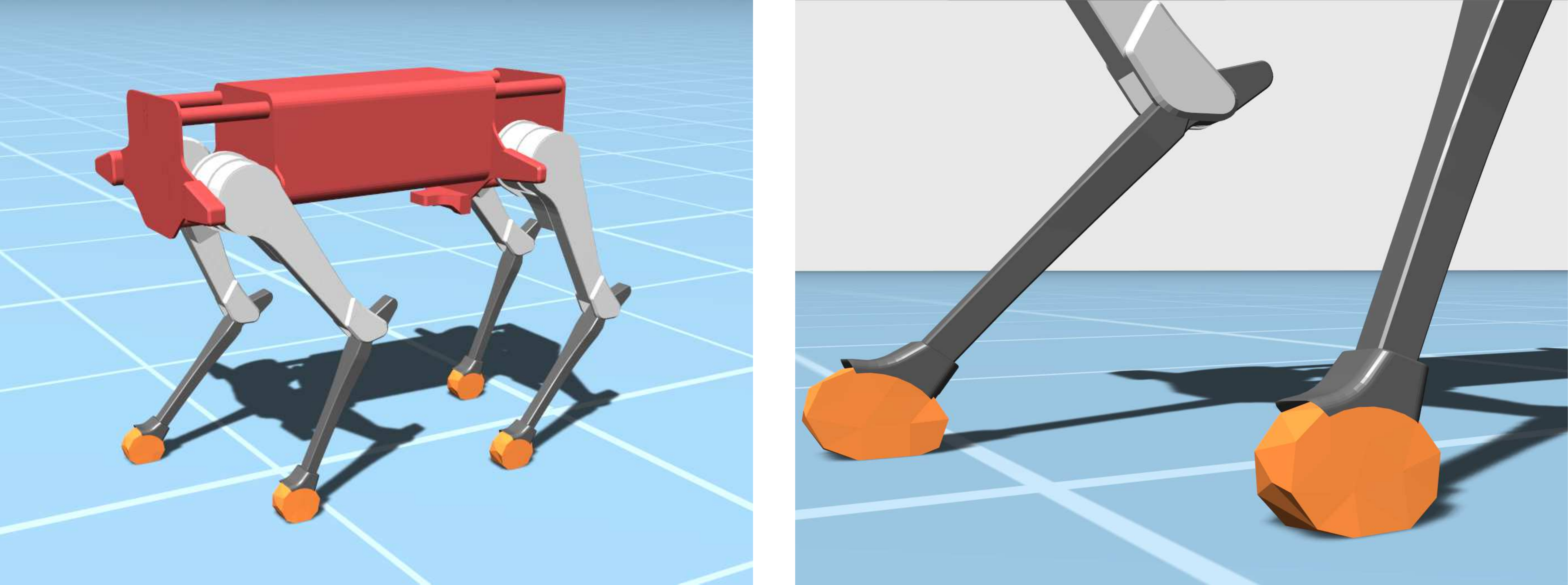}}%
    \put(0.54013486,0.34178812){\color[rgb]{0,0,0}\makebox(0,0)[lt]{\lineheight{1.25}\smash{\begin{tabular}[t]{l}(b)\end{tabular}}}}%
    \put(0.01900786,0.34209283){\color[rgb]{0,0,0}\makebox(0,0)[lt]{\lineheight{1.25}\smash{\begin{tabular}[t]{l}(a)\end{tabular}}}}%
  \end{picture}%
\endgroup%

\caption{Coupling soft and rigid bodies allows us to equip this compliant robot with soft feet. We can perform trajectory optimization on the entire robot to find control inputs that account for the compliant motors, as well as the deformable end-effectors.} \vspace{-8pt}
\label{fg:laikago-soft}
\end{figure}
\rev{Finally, we combine rigid and soft objects and equip the compliant robot with deformable feet modelled as finite element meshes, Fig.~\ref{fg:laikago-soft}. Again, the nominal motion trajectory synthesized with an idealized robot model does not carry over to the compliant robot with soft feet, as shown in the accompanying video. Using our differentiable simulator, we apply trajectory optimization to the full coupled multi-body dynamics and attain optimized controls such that the robot reaches the target distance travelled. We refer to Table~\ref{tb:ResultsRigid} for details on simulation parameters and runtime.}{}


\begin{table*}[htb]
  \centering
  \caption{Overview of material parameters, as well as simulation and optimization performance on our deformable (FEM) examples. Timings obtained on a $4\times 3.5$~GHz CPU with $16$~GB RAM. Columns: Young's modulus $E$, viscosity $\nu$ (* indicates BDF1 time integration), mass density $\rho$, coefficient of friction $c_f$ (values in braces refer to walls), number of elements in the mesh $m$, runtime of a single simulation $t_\text{sim}$, runtime for the entire optimization $t_\text{opt}$ (in seconds or h:mm:ss), number of individual simulation runs during optimization $i_\text{opt}$.
  For parameter estimation examples (Figs.~\ref{fg:SphereMarkers} and \ref{fg:CubeMarkers}, the first row gives the initial guess material parameters (in italics) with timings for optimization of initial conditions (braces) and material parameters of a single run, while the second row gives final material parameters with timings for optimization of initial conditions on a different recorded trajectory. }
\begin{tabular}{rcccccccc}
\toprule
Figure & $E$ [Pa] & $\nu$ [Pa s] & $\rho$ [kg/m$^3$] & $c_f$ [1] & $m$   & $t_\text{sim}$ [s] & $t_\text{opt}$ & $i_\text{opt}$ \\
\midrule
\ref{fg:CylinderDrop} lin. & 2.1E+03 & 0 *   & 150   & 0.4   & 879   & 26.24 &       &  \\
\ref{fg:CylinderDrop} tanh & 2.1E+03 & 0 *   & 150   & 0.4   & 879   & 30.86 &       &  \\
\ref{fg:CylinderDrop} hyb. & 2.1E+03 & 0 *   & 150   & 0.4   & 879   & 31.2  &       &  \\
\ref{fg:CylinderDrop} SQP & 2.1E+03 & 0 *   & 150   & 0.4   & 879   & 66.78 &       &  \\
\ref{fg:ToriBall-conv} lin. & 2.1E+04 & 0.1   & 150   & 0.4 (0.8) & 1930  & 119.19 &       &  \\
\ref{fg:ToriBall-conv} tanh & 2.1E+04 & 0.1   & 150   & 0.4 (0.8) & 1930  & 178.5 &       &  \\
\ref{fg:ToriBall-conv} hyb. & 2.1E+04 & 0.1   & 150   & 0.4 (0.8) & 1930  & 180.75 &       &  \\
\ref{fg:SphereToPointAndLine}a (0) & 2.0E+04 & 0.0125 & 150   & 0.4   & 451   & 6.92  & 27.33 & 4 \\
\ref{fg:SphereToPointAndLine}a (1) & 2.0E+04 & 0.0125 & 150   & 0.4   & 451   & 8.65  & 222.54 & 24 \\
\ref{fg:SphereToPointAndLine}a (2) & 2.0E+04 & 0.0125 & 150   & 0.4   & 451   & 8.62  & 265.21 & 28 \\
\ref{fg:SphereToPointAndLine}b & 2.0E+04 & 0.0125 & 150   & 0.4   & 451   & 8.21  & 01:01:11 & 453 \\
\ref{fg:SphereAndBunnyThrow}b & 2.4E+04 & 0 *   & 90    & 0.2 (0.4) & 2729  & 71.28 & 05:59:32 & 322 \\
\ref{fg:SphereMarkers} \textit{(initial mat. param.)} & \textit{2.1E+05} & \textit{0.025} & 86    & \textit{0.4} & 1351  & (4.13) 38.35 & 05:08:42 & (104) 586 \\
Second motion (video) & 3.3E+04 & 6.225 & 86    & 0.47  & 1351  & 13.38 & 158.04 & 105 \\
\ref{fg:CubeMarkers} \textit{(initial mat. param.)} & \textit{2.0E+04} & \textit{0.525} & 77    & \textit{0.4} & 1501  & (3.08) 421.70 & 06:58:22.30 & (150) 350 \\
Second motion (video) & \textit{1.5E+04} & 0.7183 & 77    & 1.16  & 1501  & 1.89  & 430.35 & 112 \\
\bottomrule
\end{tabular}%

  \label{tb:ResultsDeformable}
\end{table*}

\begin{table*}[htb]
  \centering
  \caption{Overview of simulation parameters and performance on our trajectory optimization examples. 
  Columns: number of degrees of freedom $|\mbf{q}|$, number of parameters per time step $|\mbf{p}|/n_t$, number of time steps $n_t$, time step size $\Delta t$, contact stiffness $k_n$, contact damping $k_d$, stiffness of constraints $k$ with stiffness of compliant motors (braces), motor damping $k_{md}$, coefficient of friction $c_f$, runtime for a single simulation $t_\text{sim}$, runtime for the entire optimization $t_\text{opt}$, number of optimization iterations $i_\text{opt}$.}
  \begin{tabular}{rccccccccccccc}
\toprule
Example & Figure & $|\mbf{q}|$ & $|\mbf{p}|/n_t$ & $n_t$ & $\Delta t$ [s] & $k_n$ & $k_d$ & $k$ & $k_{md}$ & $c_f$ [1] & $t_\text{sim}$ [s] & $t_\text{opt}$ & $i_\text{opt}$ \\
\midrule
Cube dragging        & video           & 12 & 6   & 60   & 1/60 & 200 & 1e-3 & 5        &    & 0.5 & 0.044 &  00:00:03.339 & 20  \\
Mass-spring flipping & \ref{fg:teaser} & 75 & 6   & 120  & 1/60 & 1e3 &  & 1000     &    & 0.5 & 0.109 & 00:00:54.419 & 25  \\
Compliant robot & \ref{fg:teaser}            & 78  & 12 & 192 & 1/60 & 1e5 & 0.1 & 5e5 (1e5) & 10 & 0.8 & 0.550 & 00:19:21.298  & 60  \\
Compliant robot, soft feet & \ref{fg:laikago-soft} & 414 & 12 & 144 & 1/60 & 1e5 &  & 5e5 (1e5) & 10 & 0.8 & 1.552 & 00:27:54.816 & 17  \\
\bottomrule
\end{tabular}%
  \label{tb:ResultsRigid}
\end{table*}

\section{Discussion}


We present an analytically differentiable dynamics solver that handles frictional contact for soft and rigid bodies. A key aspect of our approach is the use of a soft constraint formulation of frictional contact, which enables our simulation model straightforward to differentiate.
\rev{
Our results show that penalty-based contact models, especially in the normal component, are sufficiently accurate when combined with implicit time integration, 
and also enable tunable, sufficiently smooth contact treatment for gradient-based optimization.
We also analyse the effects of penalties against hard constraints with respect to static and dynamic friction.
For dynamic motion, where static friction persists only for short contact durations, penalty-based methods perform adequately and improve the performance of optimization methods built upon these simulations.
When persistent static friction is necessary, our hybrid method adds the corresponding equality constraints, but maintains soft contacts (rather than adding inequality constraints) in the normal component, and therefore still fits into our differentiable simulation framework.


Our optimization examples show that using this framework, gradient-based optimization methods greatly outperform sampling-based methods such as CMA-ES.
We demonstrate the effectiveness of our approach on a wide range of applications, including parameter estimation, robotic locomotion and manipulation tasks, as well as learning-based control.
}{Our simulation and real-world experiments showcase the promise of our model for a wide range of applications that include parameter estimation for real-world deformable objects, robotic locomotion and manipulation tasks, and learning-based control.}


All of our examples assume that contact forces obey the isotropic Coulomb model.
While this assumption is valid for a large class of surfaces, anisotropy is an important characteristic of various physical systems such as textiles \cite{Pabst2009}.
Furthermore, in our modelling, we assume a functional representation of the distance metric between different bodies to be readily available, and to be sufficiently smooth.
For complex contact scenarios where the geometric representations of the bodies involved in collisions are high resolution and highly non-convex, more elaborate collision handling methods are required~\cite{Allard2010}. 

Our experiment on learning-based control is an initial investigation into ways of combining differentiable simulators with machine learning techniques.
In the future, we see great promise in leveraging this concept in the context of deep reinforcement learning.
By eliminating the need for random exploration, for example, the analytic gradient information that our framework provides is likely to improve sample efficiency.
\rev{
Similarly, our experiments on performing throwing motions on a real-world robot demonstrate that our simulation result translate to the physical specimens.
However, there are still numerous sources of error such as aligning the robot in its environment, unwanted movements of the robot's base, and latency of the hardware controller, which require further investigation.
}{}

\rev{
In summary, our experiments demonstrate that our system enables efficient inverse problem solving for various applications in graphics and robotics.
For many of these applications, soft constraints with linear penalty forces, combined with implicit integration, lead to physically meaningful and analytically differentiable simulations.
Furthermore, we explore options for smoother friction forces, which helps reduce noise in the objective function, as well as equality constraints for static friction in cases where physical accuracy is key.
In either situation, a soft contact in the normal component enables differentiability and gradient-based optimization.
}{For trajectory optimization tasks and to robustly estimate the mechanical properties of real-world objects, continuation approaches that adaptively modulate contact stiffness have proven useful in our experiments.
}
In the future, we plan to further investigate simulation-driven optimization methods in the context of robotics and control of highly complex multi-body systems that combine rigid and flexible elements.



\bibliographystyle{ACM-Reference-Format}
\bibliography{references}

\end{document}